\documentclass[runningheads]{llncs}
\usepackage[T1]{fontenc}
\usepackage{graphicx}
\usepackage[noadjust,nospace]{cite}

\usepackage{graphics} 
\usepackage{amsmath} 
\usepackage{amssymb}  
\usepackage{mathtools}

\usepackage{bbm}
\usepackage[ruled,vlined,linesnumbered]{algorithm2e}
\newcommand\given[1][]{\:#1\vert\:}

\newtheorem{assumptionx}{Assumption}

\DeclareMathOperator*{\argmax}{arg\,max}

\DeclareMathOperator*{\spacer}{\ \vert \ }
\DeclareMathOperator*{\MJLS}{\mathfrak{J}}
\DeclareMathOperator*{\LTI}{\mathfrak{L}}
\DeclareMathOperator*{\trueP}{P^\star}
\DeclareMathOperator*{\truePtimes}{P^\star_\times}
\DeclareMathOperator*{\truePz}{P^\star_z}
\DeclareMathOperator*{\truePall}{P^\star_{\forall z}}

\usepackage{hyperref}
\usepackage{cleveref}
\usepackage{wrapfig}

\crefname{equation}{}{}
\crefname{pluralequation}{}{}

\crefname{algorithm}{Algorithm}{Algorithm}

\crefname{figure}{Fig.}{Figs.}
\crefname{pluralfigure}{Figs.}{Figs.}

\crefname{section}{Sect.}{Sects.}
\crefname{pluralsection}{Sects.}{Sects.}

\crefname{table}{Table}{Table}
\crefname{pluraltable}{Tables}{Tables}

\crefname{definition}{Def.}{Def.}
\crefname{pluraldefinition}{Defs.}{Defs.}

\crefname{theorem}{Theorem}{Theorems}
\crefname{pluraltheorem}{Theorems}{Theorems}

\crefname{lemma}{Lemma}{Lemmas}
\crefname{plurallemma}{Lemmas}{Lemmas}

\crefname{example}{Example}{Example}
\crefname{pluralexample}{Examples}{Examples}

\crefname{problem}{Problem}{Problem}
\crefname{pluralproblem}{Problems}{Problems}

\crefname{assumption}{Assumption}{Assumption}
\crefname{pluralassumption}{Assumptions}{Assumptions}

\crefname{assumptionx}{Assumption}{Assumption}
\crefname{pluralassumptionx}{Assumptions}{Assumptions}

\crefname{remark}{Remark}{Remark}
\crefname{pluralremark}{Remarks}{Remarks}

\crefname{appendix}{}{}
\crefname{pluralappendix}{}{}

\usepackage{tikz}
\usetikzlibrary{shapes.geometric, arrows.meta, positioning}
\tikzstyle{startstop} = [rectangle, rounded corners, minimum width=4cm, minimum height=0.1cm,text centered, draw=black, fill=pink, text width=5cm]
\tikzstyle{IO} = [ellipse, minimum width=2.5cm, minimum height=0.1cm,text centered, draw=black, text width=3.5cm]
\tikzstyle{arrow} = [thick,->,>={Stealth}]
\tikzset{font={\fontsize{10pt}{12}\selectfont}}

\usepackage{empheq}

\usepackage{microtype}
\usepackage[subtle]{savetrees}

\usepackage{color}

\begin{document}

\title{Formal Controller Synthesis for Markov Jump Linear Systems with Uncertain Dynamics \thanks{This work was supported by funding from the EPSRC AIMS CDT EP/S024050/1, and by NWO grant NWA.1160.18.238 (PrimaVera).}}
\titlerunning{Formal Controller Synthesis for MJLSs with Uncertain Dynamics}
%
\author{Luke Rickard\inst{1}\orcidID{0000-0002-9192-9186} \and
Thom Badings\inst{2}\orcidID{0000-0002-5235-1967} \and 
Licio Romao\inst{1}\orcidID{0000-0002-5716-6162} \and
Alessandro Abate \inst{1}\orcidID{0000-0002-5627-9093}
}
\authorrunning{L. Rickard et al.}
%
\institute{University of Oxford, Oxford, UK \and
Radboud University, Nijmegen, the Netherlands}
\maketitle              
\begin{abstract}
Automated synthesis of provably correct controllers for cyber-physical systems is crucial for deployment in safety-critical scenarios. 
However, hybrid features and stochastic or unknown behaviours make this problem challenging. 
We propose a method for synthesising controllers for Markov jump linear systems (MJLSs), a class of discrete-time models for cyber-physical systems, so that they certifiably satisfy probabilistic computation tree logic (PCTL) formulae. 
An MJLS consists of a finite set of stochastic linear dynamics and discrete jumps between these dynamics that are governed by a Markov decision process (MDP).
We consider the cases where the transition probabilities of this MDP are either known up to an interval or completely unknown.
Our approach is based on a finite-state abstraction that captures both the discrete (mode-jumping) and continuous (stochastic linear) behaviour of the MJLS.
We formalise this abstraction as an interval MDP (iMDP) for which we compute intervals of transition probabilities using sampling techniques from the so-called `scenario approach', resulting in a probabilistically sound approximation.
We apply our method to multiple realistic benchmark problems, in particular, a temperature control and an aerial vehicle delivery problem.
\end{abstract}

\section{Introduction}

In a world where autonomous cyber-physical systems are increasingly deployed in safety-critical settings, it is important to develop methods for certifiable control of these systems~\cite{knightSafetyCriticalSystems2002}.
Cyber-physical systems are characterised by the coupling of digital (discrete) computation with physical (continuous) dynamical components. 
This results in a \emph{hybrid system}, endowed with different discrete modes of operation, each of which is characterised by its own continuous dynamics~\cite{LSAZ21}. 
Ensuring that these hybrid systems meet complex and rich formal specifications when controlled is an important yet challenging goal.

\paragraph{Formal controller synthesis}
Often, these specifications cannot be expressed as classical control-theoretic objectives, which by and large relate to stability and convergence, or invariance and robustness~\cite{belta2017formal}.
Instead, these requirements can be expressed in a temporal logic, which is a rich language for specifying the desired behaviour of dynamical systems~\cite{platzerLogicsDynamicalSystems2012}.
In particular, probabilistic computation tree logic (PCTL,~\cite{hanssonLogicReasoningTime1994}) is widely used to define temporal requirements on the behaviour of probabilistic systems. 
For example, in a building temperature control problem, a PCTL formula can specify that, with at least 75\% probability, the temperature must stay within the range $22-23^\circ$C for $10$ minutes.
Leveraging probabilistic verification tools~\cite{baierPrinciplesModelChecking2008a}, it is of interest to synthesise a controller that ensures the satisfaction of such a PCTL formula for the model under study~\cite{hahnSynthesisPCTLParametric2011}. 

\paragraph{Markov jump linear systems}
Markov jump linear systems (MJLSs)~\cite{docostaDiscreteTimeMarkovJump2005} are a well-known class of stochastic, hybrid models suitable for capturing the behaviour of cyber-physical systems~\cite{LSAZ21}. 
An MJLS consists of a finite set of linear dynamics (also called \emph{operational modes}), where jumps between these modes are governed by a Markov chain (MC) or, if jumping between the modes can be controlled, by a Markov Decision Process (MDP). 
Despite each mode having linear (though possibly stochastic) dynamics, the overall dynamics are non-linear due to the jumping between modes.
MJLSs have been used to model, among other things, networked control systems, where the different operation modes relate to specific packet losses or to distinct discrete configurations~\cite{HNX07,MPPdO18}. 

\paragraph{Uncertainty in MJLSs} 
We consider a rich class of discrete-time MJLSs with two sources of uncertainty.
First, the continuous dynamics in each mode are affected by an additive stochastic process noise, e.g., due to inaccurate modelling or wind gusts affecting a drone~\cite{blackmoreProbabilisticParticleControlApproximation2010a}. 
We only assume sampling-access to the noise, rather than full knowledge of its probability distribution, allowing us to provide probably approximately correct (PAC) guarantees on the behaviour of the MJLS.
Second, similar to~\cite{MPPdO18}, we assume that the transition probabilities of the Markov jump process are not precisely known.
However, unlike \cite{MPPdO18}, we consider two different semantics for this uncertainty: either (1) transition probabilities between modes are given by intervals; or (2) these probabilities are not known at all \cite{jiangTraverseAlgorithmApproach2022,liRobustL2Filtering2013}. 
More details on the considered model are in \cref{sec:background}.

\paragraph{Problem statement} 
Several MJLS control problems have been studied, such as stability~\cite{boukasStabilityDiscretetimeLinear1995,zhangStabilityStabilizationMarkovian2009}, $H_\infty$-controller design~\cite{dFGVO00,CFT15,GG17,CGG19}, and optimal control~\cite{CF95,HSF06}.
However, limited research has been done for more complex tasks expressed in, for example, PCTL.
In this paper, we thus solve the following problem.
Given an MJLS subject to uncertainty in both its continuous dynamics (via additive noise of an unknown distribution) and its discrete behaviour (uncertain Markov jumps), compute a provably correct controller that satisfies a given PCTL formula. 

\paragraph{Abstractions of MJLSs}
We develop a new technique for abstracting MJLSs by extending methods introduced for linear non-hybrid systems in~\cite{badingsSamplingBasedRobustControl2021}.
In line with~\cite{badingsSamplingBasedRobustControl2021}, we capture the stochastic noise affecting the continuous dynamics by means of transition probability \textit{intervals} between the discrete states of the abstraction. 
We compute these intervals using sampling techniques from the \emph{scenario approach}~\cite{campiScenarioApproachSystems2009} and leverage the tighter theoretical bounds developed in \cite{romaoExactFeasibilityConvex2023a}. 
We thus formalise the resulting abstract model as an interval MDP (iMDP), which is an MDP with transition probabilities given as intervals~\cite{DBLP:journals/ai/GivanLD00}.
Different from \cite{badingsSamplingBasedRobustControl2021}, we also newly capture the discrete mode jumps in the abstract iMDP.

\paragraph{Controller synthesis}
We use the state-of-the-art verification tool PRISM~\cite{kwiatkowskaPRISMVerificationProbabilistic2011} to synthesise a policy on the abstract iMDP that satisfies a given PCTL specification. 
Leveraging results from the scenario approach, we refine this policy into a controller for the MJLS with PAC guarantees on the satisfaction of the specification.

\paragraph{Contributions} 
Our main contribution is a framework to synthesise provably-correct controllers for discrete-time MJLSs given general PCTL specifications, based on iMDP abstractions of the MJLSs. 
Previous work in this area has been limited to linear time-invariant dynamics, and to simpler reach-avoid specifications~\cite{badingsSamplingBasedRobustControl2021}.
We thus extend earlier techniques by developing new methods for a broader class of hybrid models (MJLSs) and for general PCTL formulae. 
In line with previous work, we propose a semi-algorithm based on iterative refinements of our model, meaning that a synthesised controller will satisfy the required formula, but the inability to find such a controller does not imply the non-existence of one.
Technically, we newly show how to capture both the continuous and discrete dynamics of the MJLS in the abstract iMDP model.
In particular, our methods are applicable to MJLSs where the stochastic noise in the continuous dynamics and that in the transition probabilities of the Markov jump process are unknown.

\subsection*{Related Work}
Techniques for providing safety guarantees for dynamical systems can largely be split into two approaches, respectively called \emph{abstraction-free} and \emph{abstraction-based} \cite{LSAZ21}.

Abstraction-free methods derive safety guarantees without the need to create simpler abstract models. 
For example, barrier functions \cite{lindemannLearningHybridControl2020,nejatiCompositionalConstructionControl2022,robeyLearningRobustHybrid2021} can be used to certify the existence of control inputs that keep the system within safe states.
Another approach is that of (probabilistic) reachability computation \cite{abateProbabilisticReachabilitySafety2008,moggiSafeRobustReachability2018}, where the goal is to evaluate if the system will reach a certain state over a given horizon.

Abstraction-based methods \cite{Tab09,belta2017formal} analyse a simpler model of the system, formally shown to be related to the concrete model, and thus allow to transfer the obtained results (safety guarantees, or synthesised policies) back to the original model. 
Various approaches exist for creating abstractions of different forms, including the celebrated counterexample-guided abstraction/refinement approach  \cite{clarkeVerificationHybridSystems2003} and, relevant for this work, a few involve abstractions as Markov models  \cite{abateMarkovSetChainsAbstractions2008, abateProbabilisticReachabilitySafety2008,EZSA13,cCLLAKC19,badingsRobustControlDynamical2023,Badings2022ProbabilitiesEnough_AAAI}.

Related to the approaches detailed above is robust control, where the goal is to compute a controller that achieves some task while being robust against disturbances.
Robust control techniques for MJLSs have been studied in \cite{benbrahimRobustControlConstraints2016,caiRobustModelPredictive2019,tianRobustControlMarkovian2013}.
\section{Foundations and Problem Statement}
\label{sec:background}

\subsection{Markov Decision Processes}
A Markov decision process (MDP) is a tuple $\mathcal{M}=(\mathcal{S},\mathcal{A},s_I,P)$ where $\mathcal{S}$ is a finite set of states, $\mathcal{A}$ is a finite set of actions, $s_I \in \mathcal{S}$ is the initial state, and $P \colon \mathcal{S} \times \mathcal{A} \rightharpoonup \mathit{Dist}(\mathcal{S})$ is a (partial) probabilistic transition function, with $\mathit{Dist}(\mathcal{S})$ the set of all probability distributions over $\mathcal{S}$~\cite{baierPrinciplesModelChecking2008a}.
We call a tuple $(s,a,s')$ with probability $P(s,a)(s') > 0$ a \emph{transition}.
We write $\mathcal{A}(s) \subseteq \mathcal{A}$ for the actions enabled in state $s$. 
A Markov chain (MC) is an MDP such that $|\mathcal{A}(s)| = 1, \,\forall s \in S$.
We consider time-dependent deterministic (or pure) policies, $\pi \colon \mathcal{S} \times \mathbb{N} \to \mathcal{A}$, which map states $s \in \mathcal{S}$ and time steps $k \in \mathbb{N}$, to actions $a \in \mathcal{A}(s)$.
The set of all policies for MDP $\mathcal{M}$ is denoted by $\Pi_\mathcal{M}$.

Interval Markov decision processes (iMDPs) extend regular MDPs with uncertain transition probabilities~\cite{DBLP:journals/ai/GivanLD00}.
An iMDP is a tuple $\mathcal{M}_\mathbb{I} = (\mathcal{S},\mathcal{A},s_I,\mathcal{P})$, where the states and actions are defined as for MDPs, and $\mathcal{P} \colon \mathcal{S} \times \mathcal{A} \rightharpoonup 2^{\mathit{Dist}(\mathcal{S})}$ maps states and actions to a set of distributions over successor states.
Specifically, each $\mathcal{P}(s,a)(s')$ is an \emph{interval} of the form $[\underline{p},\overline{p}]$, with $\underline{p}, \overline{p} \in (0,1], \underline{p} \leq \overline{p}$.
Intuitively, an iMDP encompasses a set of MDPs differing only in their transition probabilities: fixing an allowable probability distribution in the set $\mathcal{P}(s,a)$ for every state-action pair $(s,a)$ (denoted $P \in \mathcal{P}$ for brevity) results in an MDP, denoted by $\mathcal{M}_\mathbb{I}^P$.

\subsection{Markov Jump Linear Systems}

Let $\mathcal{Z} = \{z_1,\dots,z_N\}$ be a finite set of discrete modes. 
Consider the collections of matrices $A=(A_1,\dots,A_N)$, $A_i \in \mathbb{R}^{n\times n}$, and $B = (B_1,\dots,B_N)$, $B_i \in \mathbb{R}^{n \times m}$; and of vectors $q=(q_1,\dots,q_N), q_i \in \mathbb{R}^n$. 
A discrete-time MJLS model $\MJLS$ comprises continuous and discrete dynamics. 
Each triple $(A_i, B_i, q_i)$ defines a linear dynamical system, with discrete-time dynamics in \eqref{eq:LinearSystem:a}. 
The discrete jumps between the $N$ modes in $\mathcal{Z}$ are governed by an MDP $(\mathcal{Z},\mathcal{B},z_I, T)$ with \emph{switching actions}
$\mathcal{B}=\{1,\ldots,M\}$, and \emph{mode switch} transition function $T \colon \mathcal{Z} \times \mathcal{B} \rightharpoonup \mathit{Dist}(\mathcal{Z})$.
At any time $k \in \mathbb N$,  
we denote the continuous state by $x(k) \in \mathcal{X} \subseteq \mathbb{R}^n$ ($\mathcal{X}$ bounded), and the discrete mode by $z(k) \in \mathcal{Z}$. 
Given initial state $x(0) \in \mathcal{X}, z(0) \in \mathcal{Z}$, the (hybrid) state $(x,z)$ is computed as
\begin{subequations}
    \label{eq:LinearSystem}
    \begin{empheq}[left=\MJLS \colon \empheqlbrace]{align}
        x(k+1) &= A_{z(k)}x(k)+B_{z(k)}u(k)+q_{z(k)}+w_{z(k)}(k) 
        \label{eq:LinearSystem:a}
        \\ 
        z(k+1) &\sim T(z(k), b(k)),
        \label{eq:LinearSystem:b}
    \end{empheq}
    \label{eq:LTI_system}%
\end{subequations}%
where $u(k) \in \mathcal{U} \subseteq \mathbb{R}^m$ is the control input to the continuous dynamics, and $b(k)\in \mathcal{B}$ is the discrete (MDP) switching action.
Note, for each mode $z \in \mathcal{Z}$, the corresponding continuous dynamics are affected by an additive stochastic process noise $w_{z}$, with a (potentially) unknown distribution. 
The distribution of the noise $w_z$ is not required to be the same across different modes, but $\{w_z(k)\}_{k \in \mathbb{N}}$ must be an i.i.d. stochastic process having density with respect to the Lebesgue measure, and independent across modes. 
Importantly for our setting, the input $u$ and switch $b$ are \emph{jointly determined} by a feedback controller (namely, a policy for the MJLS) of the following form.
\begin{definition}
    A time-dependent feedback \emph{controller} $F \colon \mathcal{X} \times \mathcal{Z} \times \mathbb{N} \to \mathcal{U} \times \mathcal{B}$ is a function that maps a continuous state $x \in \mathcal{X}$, discrete mode $z \in \mathcal{Z}$, and time step $k \in \mathbb{N}$,  to a continuous control input $u \in \mathcal{U}$ and discrete switch $ b \in \mathcal{B}$.
    \label{def:ControlLaw}
\end{definition}
\begin{example}
\label{example:temp1}
Consider a temperature regulation problem inspired by~\cite{abateProbabilisticReachabilitySafety2008}, in which a portable fan heater and a portable radiator are used to heat a two-room building.
We define two modes $\mathcal{Z}=\{1,2\}$, relating to the fan heater being in room 1 or 2 respectively (and the radiator in the other room). 
Swapping the heat sources between rooms is modelled by an MDP with actions $\mathcal{B}=\{0,1\}$, relating to leaving or switching the heaters.
Each of these mode-switching actions fails with some probability. 
This problem is naturally modelled as an MJLS with the matrices
\begin{equation}
\begin{split}
    A_{\{1,2\}} &= \begin{bmatrix} 
    1-b_1-a_{12} & a_{12}\\
    a_{21} & 1-b_2-a_{21}
    \end{bmatrix}, 
    \\
    B_1 &= \begin{bmatrix} 
    k_f & 0\\
    0 & k_r
    \end{bmatrix}, \
        B_2 = \begin{bmatrix} 
    k_r & 0\\
    0 & k_f
    \end{bmatrix}, \
    q_{\{1,2\}} = \begin{bmatrix}
    b_1x_a\\
    b_2x_a
    \end{bmatrix},
\end{split}
\label{eq:example:dynamics}
\end{equation}
where the state $x = [T_1, T_2]^\top \in \mathbb{R}^2$ models the room temperatures, and the power of both heaters can be adjusted within the range $u \in [0,1]^2$ (the extrema denoting being fully on and off).
In \cref{sec:results}, we perform a numerical experiment with this MJLS.
\qed
\end{example}

\subsection{Probabilistic Computation Tree Logic}
\label{sec:PCTL}

Probabilistic computation tree logic (PCTL) depends on the following syntax~\cite{baierPrinciplesModelChecking2008a}:
\begin{equation}
\begin{aligned}
        \Phi &::= true \spacer p \spacer \neg \Phi \spacer \Phi \wedge \Phi \spacer \mathsf{P}_{\sim \lambda}(\psi)\\
        \psi &::= \Phi \mathsf{U} \Phi \spacer \Phi \mathsf{U}^{\leq K} \Phi \spacer \mathsf{X}\Phi.
\end{aligned}
\end{equation}
\noindent
Here, $\sim \in \{ <, \leq, \geq, > \}$ is a comparison operator and $\lambda \in [0,1]$ a probability threshold; PCTL formulae $\Phi$ are state formulae, which can in particular depend on path formulae $\psi$.   
Informally, the syntax consists of state labels $p \in AP$ in a set of atomic propositions $AP$, propositional operators negation $\neg$ and conjunction $\wedge$, and temporal operators until $\mathsf{U}$, bounded until $\mathsf{U}^{\leq K}$, and next $\mathsf{X}$. 
The probabilistic operator $\mathsf{P}_{\sim \lambda}(\psi)$ requires that paths generated from the initial conditions satisfy a path formula $\psi$ with total probability exceeding (or below, depending on $\sim$) some given threshold $\lambda$.

An MJLS $\MJLS$ with a controller $F$ induces a stochastic process on the hybrid state space $\mathcal{X} \times \mathcal{Z}$.
Let $L_\mathcal{J} \colon \mathcal{X} \times \mathcal{Z} \to 2^{AP}$ be a labelling from hybrid states to a subset of labels.
Recall that the noise affecting the continuous dynamics in \cref{eq:LTI_system} has density with respect to the Lebesgue measure.
We assume for each label that the set $\{ x \in \mathcal{X} \colon p \in L_\mathcal{J}(x,z), z \in \mathbb{Z} \} \subseteq \mathcal{X}$ of continuous states with label $p$ is measurable.
We follow the same semantics as used in \cite{DBLP:journals/tac/LahijanianAB15} for stochastic hybrid systems, i.e., the (initial) state $x(0),z(0)$ of an MJLS $\MJLS$ satisfies a property $\Phi = \mathsf{P}_{\sim \lambda}(\psi)$ if the probability of all paths from $x(0),z(0)$ satisfies $\sim \lambda$.
For brevity, we shall write this satisfaction relation as $\MJLS \models_F \Phi$.
All the sets of paths $(x(0), z(0)), (x(1), z(1)), \ldots$ expressed by PCTL under the above assumptions are measurable, see, e.g.,~\cite{DBLP:conf/hybrid/RamponiCSL10,TA14,LSAZ21} for details.

For an iMDP $\mathcal{M}_\mathbb{I}$, the satisfaction relation $\mathcal{M}_\mathbb{I} \models_\pi \Phi$ defines whether a PCTL formula $\Phi$ holds true, when following policy $\pi$ from the initial state(s).
Formal definitions for semantics and model checking are provided in \cite{hanssonLogicReasoningTime1994,baierPrinciplesModelChecking2008a}.
Recall from \cref{eq:opt_imdp} that for iMDPs $\mathcal{M}_\mathbb{I}$, the threshold $\sim \lambda$ must hold under the \emph{worst-case realization} of the probabilities $P \in \mathcal{P}$ in their intervals.
That is, we are interested in synthesizing an optimal policy $\pi^\star \in \Pi_{\mathcal{M}_\mathbb{I}}$ that maximises the probability of satisfying a path specification $\psi$ for the \emph{worst-case} assignment $P \in \mathcal{P}$ (which is determined by a so-called \emph{adversary}). 
In other words, we seek to solve the max-min decision problem given by
\begin{equation}
\label{eq:opt_imdp}
    \pi^\star = \argmax_{\pi \in \Pi_{\mathcal{M}_\mathbb{I}}} \ \min_{P \in \mathcal{P}} \lambda \quad \text{s.t.} \ \; \mathsf{P}_{\geq \lambda}(\mathcal{M}_\mathbb{I}^P \models_\pi \psi).
\end{equation}
It is shown by \cite{puggelliPolynomialTimeVerificationPCTL2013,cZWIA18}, and in a much more general setting by \cite{DBLP:journals/siamco/Gonzalez-TrejoHH02}, that deterministic policies suffice to obtain optimal values for iMDPs. 

\subsection{Problem Statement}

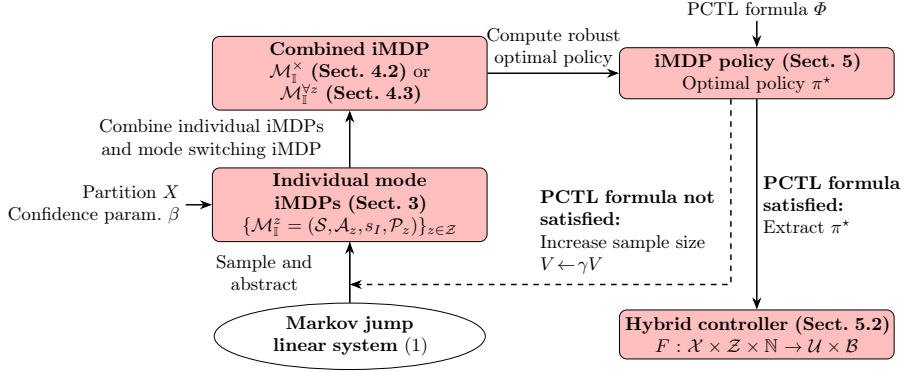
\begin{figure}[!t]
\vspace{-1.5cm}
\centering
     \scalebox{0.7}{
\begin{tikzpicture}[node distance=3cm]
    \node (root2) {};
    \node (root) [below of=root2] {};
    \node (inputs) [text width=3.3cm, align=right, below of=root, xshift=-1.5cm]  {Partition $X$ \\ Confidence param. $\beta$};
    \node (ind_imdp) [startstop, right=0.5cm of inputs] {\textbf{Individual mode iMDPs (\cref{sec:AAAI})} $\{\mathcal{M}^z_\mathbb{I}=(\mathcal{S}, \mathcal{A}_z, s_I, \mathcal{P}_z)\}_{z \in \mathcal{Z}}$};
    \node (imdp) [startstop, above of=ind_imdp, yshift=-0.5cm] {\textbf{Combined iMDP} \\ $\mathcal{M}_\mathbb{I}^\times$ \textbf{(\cref{subsec:main:AssumptionA})} or \\
    $\mathcal{M}_\mathbb{I}^{\forall z}$ \textbf{(\cref{subsec:main:AssumptionB})}};
    \node (sys) [IO, inner sep=2pt, below of=ind_imdp, yshift=0.5cm] {\textbf{Markov jump} \\ \textbf{linear system} \cref{eq:LinearSystem}};
    \node (prism) [startstop, right=2.5cm of imdp] {\textbf{iMDP policy (\cref{subsec:main:pol})}\\Optimal policy $\pi^\star$};
    \node (cont) [startstop, below=4.0cm of prism] {\textbf{Hybrid controller (\cref{subsec:main:controller})}\\ $F:\mathcal{X} \times \mathcal{Z} \times \mathbb{N} \to \mathcal{U} \times \mathcal{B}$};
    \node (PCTL) [text centered, above=0.5cm of prism, text width=3cm] {PCTL formula $\Phi$};
    
    \draw [arrow] (inputs) -- (ind_imdp);
    \draw [arrow] (PCTL) -- (prism);
    \draw [arrow] (imdp) -- node[anchor=south,text width=3cm, text centered] {Compute robust \\optimal policy} (prism);
    \draw [arrow] (sys) -- node[anchor=east,text width=3cm, text centered] {Sample and abstract} (ind_imdp);
    \draw [arrow] (ind_imdp) -- node[anchor=east,text width=5cm, text centered] {Combine individual iMDPs and mode switching iMDP} (imdp);
    \draw [arrow] (prism) -- node[anchor=west,text width=3cm, align=left] {\textbf{PCTL formula satisfied:}\\Extract $\pi^\star$} (cont);

    \draw [-] (sys) -- coordinate[pos=0.3] (sample) (ind_imdp);
    \draw [dashed, thick, , arrows = {-Stealth}] (prism.225) |- node[above=1cm,left,text width=3.5cm] {\textbf{PCTL formula not satisfied:}\\Increase sample size $V\leftarrow\gamma V$} (sample);
\end{tikzpicture}
}
    \caption{Approach for synthesising a provably-correct controller for an MJLS.}
    \label{fig:overview}
\end{figure}

We consider tasks encoded as a PCTL formula $\Phi$. 
Our goal is to find a feedback controller $F$ that satisfies $\Phi$.
As such, we solve the following problem.
\begin{problem}
    \label{prob:Formal}
    Given an MJLS $\MJLS$ as in \eqref{eq:LinearSystem} and a PCTL formula $\Phi$, find a control policy $F \colon \mathcal{X} \times \mathcal{Z} \times \mathbb{N} \to \mathcal{U} \times \mathcal{B}$, such that $\MJLS \models_F \Phi$.
\end{problem}
\noindent
In this paper, we address \cref{prob:Formal} through the lens of abstractions \cite{Tab09}, under two distinct assumptions on the mode-transition function of the MJLS. 
\begin{assumptionx}[Uncertain Markov jumps]
\label{ass:MDP}
Each transition probability of the MDP $(\mathcal{Z}, \mathcal{B}, z_I, T)$ driving the jumps across modes in $\mathcal{Z}$ is known up to a certain interval $\mathcal{T} \ni T$, i.e., the Markov jump process is an iMDP $(\mathcal{Z}, \mathcal{B}, z_I, \mathcal{T})$. 
\end{assumptionx}
\begin{assumptionx}[Unknown Markov jumps]
\label{ass:unknown}
The Markov jumps are driven by a Markov chain (MC) for which we can measure the current mode, but the transition function (and hence its underlying graph structure) is unknown. 
\end{assumptionx}
For \cref{ass:MDP}, we will exploit the transition function of the jump process to reason over the joint probability distribution over the state $x(k)$ and mode $z(k)$.
By contrast, under \cref{ass:unknown}, we do not know the distribution over successor modes $z(k+1)$, so reasoning over the joint distribution is not possible.
Instead, our goal is to attain \emph{robustness} against any mode changes that may occur.
An overview of our abstraction-based approach to solve \cref{prob:Formal} is presented in~\cref{fig:overview}. 
We note that it may occur that the PCTL formula is not satisfiable on the abstract model.
To alleviate this issue, we propose an iterative refinement of the abstraction (shown by the dashed line in \cref{fig:overview}), which we explain in more detail in \cref{subsec:main:pol}.

\section{Abstractions of Non-Hybrid Dynamical Systems}
\label{sec:AAAI}

Our abstraction procedure expands on the techniques from~\cite{badingsSamplingBasedRobustControl2021,badingsRobustControlDynamical2023} to make them applicable to hybrid (and probabilistic) models. 
We start by summarising the main contributions of these papers, while referring to~\cite{badingsSamplingBasedRobustControl2021,badingsRobustControlDynamical2023} for proofs and more details.

Consider a discrete-time linear system $\LTI$ with additive stochastic noise: 
\begin{equation}
    \label{eq:LTI}
    \LTI \colon x(k+1) = Ax(k)+Bu(k)+q+w(k),
\end{equation}
where $A \in \mathbb{R}^{n\times n}, B \in \mathbb{R}^{n \times m}, q \in \mathbb{R}^n$, and $w(k)$ defines an i.i.d. stochastic process, and $x(k) \in \mathcal{X} \subseteq \mathbb{R}^n$ and $u(k) \in \mathcal{U} \subseteq \mathbb{R}^m$ are the states and control inputs, respectively.
The distribution of the noise $w(k)$ is assumed to be unknown, but instead we have access to a set $\{\delta_1, \dots, \delta_{V}\}$ of $V$ i.i.d. samples of $w(k)$.
Note that the system in \cref{eq:LTI} reduces to an MJLS with a single mode. 
Given such a set of i.i.d. samples, the authors in~\cite{badingsRobustControlDynamical2023} show how to construct an iMDP which, with a specified confidence level, abstracts the system in \cref{eq:LTI}:
\begin{definition}[$\beta$-iMDP abstraction]
    \label{def:iMDP_abstraction}
    Choose $\beta \in (0,1)$ and let $\{\delta_1, \dots, \delta_{V}\}$ be a collection of samples from the noise distribution affecting the dynamics in \cref{eq:LTI}.
    An iMDP $\mathcal{M}_\mathbb{I}=(\mathcal{S},\mathcal{A},s_I,\mathcal{P})$ is a $\beta$-iMDP abstraction if for every PCTL formula $\Phi$ and for every policy $\pi \in \Pi_{\mathcal{M}_\mathbb{I}}$, there exists a feedback control $F \colon \mathcal{X} \times \mathbb{N} \to \mathcal{U}$ such that, for any initial condition $x(0)$, we have that
    \begin{equation}
        \mathbb{P}^V \Big\{ 
             (\mathcal{M}_\mathbb{I} \models_\pi \Phi) \implies
            (\LTI \models_F \Phi)
        \Big\} \geq 1 - \beta,%
    \end{equation}%
    where $s_I$ is the initial state of the $\beta$-iMDP associated with continuous state $x(0)$, and $\mathbb{P}^V$ is the product probability measure induced by the sample set $\{\delta_1, \ldots, \delta_V\}$. 
\end{definition}
We remark that $\mathbb{P}^V$ is the product probability measure corresponding with sampling a set $\{\delta_1, \ldots, \delta_V\}$ of $V \in \mathbb{N}$ samples of the noise $w(k$) in \cref{eq:LTI} (see, e.g.,~\cite{campiExactFeasibilityRandomized2008} for details). 
\cref{def:iMDP_abstraction} states that, with a confidence of at least $1-\beta$, the satisfaction of a formula on the abstract iMDP implies the existence of a feedback controller that allows the satisfaction of the same formula on the concrete model. 
The confidence bound accounts for the inherent statistical error caused by constructing the iMDP based on a finite set of noise samples only. 
The iMDP abstraction allows us to synthesise correct-by-design feedback controllers for continuous-state dynamical systems~\cite{DBLP:conf/cav/MazoDT10}, by utilising policies designed for a discrete-state model.
Note that \cref{def:iMDP_abstraction} applies to general PCTL formulas, while~\cite{badingsRobustControlDynamical2023} only considers reach-avoid properties (a subset of PCTL). 
\begin{definition}[Partition] 
    \label{def:partition}
    A partition $X = \{X_1,\ldots,X_p\}$ is an ordered set of subsets of $\mathcal{X}$ such that $\mathcal{X} = \bigcup_{i=1}^p X_i$, and $X_i \bigcap X_j = \emptyset, \,\, \forall i,j \in \{1,\ldots,p\}, \,\, i \neq j$.
\end{definition}
Papers \cite{badingsSamplingBasedRobustControl2021,badingsRobustControlDynamical2023} show how to generate $\beta$-iMDP abstractions by combining partitioning of the state space, backward reachability computation, and the scenario approach theory~\cite{campiExactFeasibilityRandomized2008}.
To this end, these papers create an iMDP abstraction $(\mathcal{S},\mathcal{A},s_I,\mathcal{P})$ of the continuous-state dynamics using the following procedure:
\begin{itemize}
    \item The set of states $\mathcal{S} = \{ s_1,\ldots,s_p \} \cup \{ s^\star \}$ consists of elements associated with a partition $X$ of the state space. This correspondence is given by the quotient mapping induced by the equivalence relation of the partition (see, e.g.,~\cite{Tab09}).
    \item The action space $\mathcal{A} = \{ a_1,\ldots,a_q \}$, where each action $a \in \mathcal{A}$ is associated with a target point $d \in \mathcal{X}$ in the continuous state space (a convenient choice is to define each target $d$ as the centre of an element $X_i \in X$ of the partition).
    \item To decide which actions are enabled at a given state of the abstraction, backward reachable set computations are employed. More specifically, we let
    \begin{equation}
        \label{eq:backreachset}
        \mathcal{R}^{-1}(a) = \{ x \in \mathbb{R}^n \mid d = Ax + Bu + q, \, u \in \mathcal{U} \} 
    \end{equation}
    be the backward reachable set of the target point $d$ associated with the action $a \in \mathcal{A}$.
    Action $a$ is enabled in state $s \in \mathcal{S}$ if and only if its corresponding element $X_i \in X$ is contained in $\mathcal{R}^{-1}(a)$.
    Mathematically, we have that
    \begin{equation}
        \label{eq:enabledactions}
        \mathcal{A}(s) = \big\{ 
            a \in \mathcal{A} \mid
            X_i \subseteq 
            \mathcal{R}^{-1}(a)
        \big\}.
    \end{equation}
    \item The initial state $s_I$ of the iMDP is defined by the element of the partition to which the initial state of the continuous dynamics belongs.
    \item The probability intervals $\mathcal{P}(s,a_j)(s_i)$ of the abstract iMDP can be efficiently computed using the scenario approach~\cite{campiSamplingandDiscardingApproachChanceConstrained2011a, romaoExactFeasibilityConvex2023a}, or using statistical inequalities such as Hoeffding's bound~\cite{DBLP:books/daglib/Boucheron2013}.
\end{itemize}
To show that this procedure indeed yields a $\beta$-iMDP abstraction as per \cref{def:iMDP_abstraction}, we also invoke the following key result from~\cite{badingsRobustControlDynamical2023}: 
\begin{theorem}[iMDP abstraction of stochastic linear systems~\cite{badingsRobustControlDynamical2023}]
\label{thm:AAAI}
    Let $X$ be a partition of the state space, then for any $\beta \in (0,1)$ and sample set $\{\delta_1,\ldots,\delta_V \}$, the procedure above yields a $\beta$-iMDP abstraction for the dynamics in \cref{eq:LTI}.
\end{theorem}
We provide an intuitive proof outline here, while referring to \cite{badingsRobustControlDynamical2023} for the full proof. 
Consider state $s$, with an associated continuous state $x(k)$; successor state $s'$, with associated partition $X_i$; action $a$, with an associated feedback controller $F$.
The true probability of transitioning from $s$ to $s'$, under action $a$ is defined as
\begin{equation}
    \trueP(s,a)(s') = \int_{\mathbb{R}^n}\mathbbm{1}_{X_i}\left(Ax(k)+BF(x(k),k)+q+\xi\right) \mathbb{P}_w(d\xi) ,
\end{equation}%
where $\mathbb{P}_w$ is the (in practice unknown) probability measure induced by the noise distribution, $\mathbbm{1}_{X_i}(\cdot)$ is the indicator function (which returns value 1 if its argument belongs to the set $X_i$).
Since transition probability intervals are obtained from the scenario approach theory, they contain probability $\trueP$ with confidence $\beta$: 
\begin{equation}
    \label{eq:AAAI:prob_intervals}
    \mathbb{P}^V \Big\{ 
        \trueP(s,a)(s') \in  \mathcal{P}(s,a)(s'), \forall s \in \mathcal{S}
    \Big\} \geq 1 - \frac{\beta}{|\mathcal{A}| \cdot |\mathcal{S}|}.
\end{equation}%
The generated iMDP has at most $|\mathcal{A}|\cdot |\mathcal{S}|$ unique probability intervals, because $\trueP(s,a)(s') = \trueP(s'',a)(s')$ for any $s,s'' \in \mathcal{S}$ in which $a$ is enabled.
Thus, using Boole's inequality, we have that for all probabilities $\trueP(s,a)(s')$ 
\begin{equation}
\mathbb{P}^V \Big\{ 
    \trueP(s,a)(s') \in  \mathcal{P}(s,a)(s'), \forall s, s' \in \mathcal{S}, a \in \mathcal{A}
\Big\} \geq 1 -  \beta.
\label{eq:thm1:proof1}
\end{equation}%
Let $\mathcal{M}^{\trueP}$ denote the MDP under the true transition function $\trueP$, and let $\pi \in \Pi_{\mathcal{M}^{\trueP}}$ be any policy for this MDP such that a given PCTL property $\Phi$ is satisfied on the iMDP, i.e. $\mathcal{M}^{\trueP} \models_\pi \Phi$.
Using concepts from probabilistic simulation relations~\cite{hermannsProbabilisticLogicalCharacterization2011,cCLLAKC19,LSAZ21}, it can be shown that there exists a controller $F$ 
such that $(\mathcal{M}^{\trueP} \models_\pi \Phi) \implies (\LTI \models_F \Phi)$.
Combining this with \cref{eq:thm1:proof1}, which states that $ \mathbb{P}^V \{\mathcal{M}^{\trueP} \in \mathcal{M}_\mathbb{I} \} \geq 1-\beta$, we arrive at the condition for a $\beta$-iMDP abstraction in \cref{def:iMDP_abstraction}.

\cref{thm:AAAI} can be used to synthesise \emph{provably correct} controllers for temporal logic specifications, but is limited to systems \emph{without discrete dynamics}, as for MJLSs.
In what follows, we will develop a framework to overcome this limitation. 
\section{Abstractions of Markov Jump Linear Systems}
\label{sec:main}
In this section, we present our main contributions to solving \cref{prob:Formal}.
We first explain how we use the results from \cref{sec:AAAI} to construct an abstraction for the continuous dynamics of an individual mode.
Then, we discuss how to ``combine'' abstractions across discrete modes to obtain a single iMDP abstraction.
Finally, we compute an optimal policy $\pi^\star$ on the obtained iMDP and show (using \cref{thm:sound,thm:controlled_sound}) that this policy can be refined as a controller for the hybrid dynamics.

\subsection{iMDP Abstraction for Individual Modes}
\label{sec:individual_abstraction}

\label{subsec:Main:Individual}

We construct an abstraction for each separate mode $z \in \mathcal{Z}$ of the MJLS defined by \cref{eq:LinearSystem} using the procedure that led to \cref{thm:AAAI}.
For simplicity, we consider rectangular partitions, but our methods are applicable for any partition into convex sets satisfying \cref{def:partition}, and even to distinct partitions across modes.
We then obtain a $\beta$-iMDP abstraction $\mathcal{M}_\mathbb{I}^z=(\mathcal{S},\mathcal{A}_z, s_I, \mathcal{P}_z)$ for each mode $z \in \mathcal{Z}$.

In order to reason over the \emph{hybrid} system as a whole, we now need a sound method to ``combine''  the abstractions $\mathcal{M}_\mathbb{I}^z$ for each mode $z \in Z$ into a single abstract model.
However, without careful consideration of the enabled discrete actions, the resulting model may fail to soundly abstract the overall MJLS, as different actions may be enabled in the same region of continuous states, and this would lead to spurious trajectories in the abstraction. 

To exemplify this issue, consider a specific instance of \cref{example:temp1}, in which the room without the radiator is perfectly insulated from the other. 
If we naively ``combine'' single-mode abstractions together, then we might conclude that we will be able to heat either room to any temperature, since the two modes taken individually can heat either room.
This is an example of artificial behaviour introduced in the abstraction.
In reality, we can only control one room at the same time; any actions which say otherwise will not be realisable on the concrete dynamical model. 
As our main contribution, we introduce in \cref{subsec:main:AssumptionA} an approach for combining single-mode iMDPs under \cref{ass:MDP} in a sound manner, and in \cref{subsec:main:AssumptionB} we discuss the case for \cref{ass:unknown}. 

\subsection{Abstraction Under Uncertain Markov jumps (Assumption A)}
\label{subsec:main:AssumptionA}

Under \cref{ass:MDP}, we have access to an iMDP representation $\mathcal{M}_\mathbb{I}=(\mathcal{Z},\mathcal{B},z_I,\mathcal{T})$ of the discrete-mode Markov jump process, which has modes in $\mathcal{Z}$, switching actions in $\mathcal{B}$, initial mode $z_I$, and transition probability intervals in $\mathcal{T}$. 
Let $\{\mathcal{M}^z_\mathbb{I}=(\mathcal{S},\mathcal{A}_z,s_I,\mathcal{P}_z)\}_{z\in\mathcal{Z}}$ be a set of $\beta$-iMDPs for each mode $z \in Z$, constructed as described in \cref{subsec:Main:Individual} with a confidence level of $\beta \in (0,1)$.
We assume that these $\beta$-iMDPs have a common state space $\mathcal{S}$, and an overall action space $\mathcal{A}$. 
We also allow for a mode-dependent set of enabled actions; and use the notation $\mathcal{A}_z(s)$ to define actions enabled at a state $s$, in mode $z$. 

To combine these modes, we use a product construction, similar to methods for constructing product automata \cite{DBLP:series/eatcs/Gecseg86}.
We define our product construction among $\mathcal{M}_\mathbb{I}$ and $\{\mathcal{M}^z_\mathbb{I}\}_{z\in\mathcal{Z}}$.
The \emph{joint state/action} space of the product are the sets $\mathcal{Z} \times \mathcal{S}$ and $ \mathcal{B} \times \mathcal{A}$. 
At a particular joint state $(z, s)$, we define the set of enabled actions $\mathcal{A}(z,s) =  \mathcal{B}(z) \times \mathcal{A}_z(s)$ as the product between the actions enabled at a particular mode, and the switches allowed in the corresponding state of the discrete iMDP. 
Thus, an action in the product iMDP corresponds with executing both an action in $\mathcal{A}_z$ (for the current mode $z$) and a discrete mode switching action in $\mathcal{B}$.
The overall product iMDP under \cref{ass:MDP} is defined as follows:

\begin{definition}[Product iMDP with mode switch control]
\label[definition]{def:prod_imdp}
Let $\{\mathcal{M}^z_\mathbb{I}=(\mathcal{S},\mathcal{A}_z,s_I,\mathcal{P}_z)\}_{z\in\mathcal{Z}}$ be a set of $\beta$-iMDP abstractions for each mode $z \in Z$, and let $\mathcal{M}_\mathbb{I}=(\mathcal{Z},\mathcal{B},z_I,\mathcal{T})$ be an iMDP for the Markov jump process.
Then, the product iMDP $\mathcal{M}_\mathbb{I}^\times=(S_\times, A_\times, s^I_\times, \mathcal{P}_\times )$ is defined with%
 \begin{itemize}
     \item Joint state space $S_\times=\mathcal{Z} \times \mathcal{S}$;
     \item Joint action space $\mathcal{A}_\times = \mathcal{B} \times \mathcal{A}$, with enabled actions $\mathcal{A}(z,s)$ in state $(z,s)$;
     \item Initial joint state $s^I_\times=(z_I, s_I)$;
     \item For each $(z,s), (z', s') \in \mathcal{Z} \times \mathcal{S}$ and $(b,a) \in \mathcal{A}(z,s)$, the probability interval 
        \begin{equation}
        \label{eq:product_imdp_interval}
            \begin{aligned}
                \mathcal{P}_\times\big( (z,s),(b,a) \big)&\big( (z',s') \big) = \\
            	&[\underline{t}(z,b)(z') \cdot \underline{p_{z}}(s,a)(s'), \;
            	 \overline{t}(z,b)(z') \cdot \overline{p_{z}}(s,a)(s')] .
            \end{aligned}
        \end{equation}
 \end{itemize}
    Here $\underline{p_z}(s,a)(s')$ and $\overline{p_{z}}(s,a)(s')$ are, respectively, the lower and upper bound state transition probability of $\beta$-iMDP $\mathcal{M}^z_\mathbb{I}$ for mode $z \in \mathcal{Z}$, and $[\underline{t}(z,b)(z'), \, \overline{t}(z,b)(z')]$ are the intervals in the transition function $\mathcal{T}$ of the jump process iMDP.
\end{definition}

By construction, the product iMDP merges the individual mode abstractions and the mode-switching iMDP in a sound manner, thus avoiding the issues with spurious actions described in \cref{subsec:Main:Individual}. 
The product iMDP depends on $NV$ samples ($N$ sets of $V$ samples, one for each mode), hence the abstraction is a random variable on the space $NV$.
The $\mathbb{P}^{NV}$ appearing in these theorems denotes the product measure $\mathbb{P}_{z_1}^V \otimes \mathbb{P}_{z_2}^V \cdots \mathbb{P}_{z_N}^V$ (note that the noise distribution can differ between modes).
We extend \cref{thm:AAAI} to the product iMDP as follows. 

\begin{theorem}[iMDP abstraction of controlled MJLS]
\label{thm:controlled_sound}
        The product iMDP defined by \cref{def:prod_imdp} is a $\beta'$-iMDP abstraction with confidence $\beta' = \beta \cdot |\mathcal{Z}|$ for the MJLS in \cref{eq:LinearSystem}, which captures the mode switching iMDP $\mathcal{M}_\mathbb{I}$.
        In particular,
        \begin{equation}
            \mathbb{P}^{NV} \Big\{ 
                 (\mathcal{M}^\times_\mathbb{I} \models_\pi \Phi) \implies
                (\MJLS \models_F \Phi)
            \Big\} \geq 1 - \beta'.
        \end{equation}%
\end{theorem}

We provide an outline of the proof here, whilst for a detailed proof we refer to \cref{app:proofs}.
The key observation is that the product iMDP is defined as the product between $|Z|$ $\beta$-iMDPs (having intervals that are ``correct'' with probability at least $1-\frac{\beta}{|\mathcal{A}| \cdot |\mathcal{S}|}$, cf. \cref{eq:AAAI:prob_intervals}) and the mode switching iMDP (which is ``correct'' with probability one).
These $|\mathcal{Z}|$ individual-mode iMDPs have $|\mathcal{A}| \cdot |\mathcal{S}| \cdot |\mathcal{Z}|$ unique intervals in total.
Thus, the probability for all intervals to be correct (and thus for the product iMDP to be sound) is at least $1-\frac{\beta \cdot |\mathcal{A}| \cdot |\mathcal{S}| \cdot |\mathcal{Z}|}{|\mathcal{A}| \cdot |\mathcal{S}|} = 1-\beta'$.
Finally, analogously to \cref{thm:AAAI}, the iMDP is a probabilistic simulation relation~\cite{hermannsProbabilisticLogicalCharacterization2011}, such that the satisfaction of general PCTL formulae in the discrete abstraction guarantees the satisfaction of the same formulae in the concrete MJLS system.

\subsection{Abstraction Under Unknown Markov jumps (Assumption B)}
\label{subsec:main:AssumptionB}

Under \cref{ass:unknown}, the mode transition probabilities are now completely unknown. 
Thus, in contrast with \cref{subsec:main:AssumptionA}, we generate an abstraction that is robust to any mode we may be in.

\paragraph{Robustifying enabled actions}
First, we modify the computation of the backward reachable set in \cref{eq:backreachset} to introduce a backward reachable set across all possible modes (i.e., the set that can reach $d$ regardless of which mode we are in -- note the universal quantification $\forall z \in \mathcal{Z}$ in the following equation): 
\begin{equation}
\label{eq:all_back}
\begin{aligned}
    \mathcal{G}^{-1}(d) &=\{ x \in \mathbb{R}^n \given d = A_zx+B_zu+q_z,
    u \in \mathcal{U},  \forall z \in \mathcal{Z}\}\\
    &=\bigcap_{z\in \mathcal{Z}}\{ x \in \mathbb{R}^n \given d = A_z x+B_zu+q_z,
    u \in \mathcal{U}\}
    =\bigcap_{z\in \mathcal{Z}} \mathcal{R}^{-1}_z(d),
    \end{aligned}
\end{equation}
where $\mathcal{R}_z^{-1}(d)$ is the backward reachable set for mode $z \in \mathcal{Z}$, as defined in \cref{eq:backreachset}.

Similar to \cref{sec:AAAI}, we use backward reachable set computation to define the set of enabled actions, now denoted by $\mathcal{A}_{\forall z}$, in the iMDP.
Indeed, for a given partition of the state space $X = \{X_1,\dots,X_p\}$, an action $a$ is enabled at a state $s$ if the backward reachable set $\mathcal{G}^{-1}(d)$ defined in \cref{eq:all_back} contains the corresponding element $X_i$ of the partition $X$, i.e., $a$ is enabled if $X_i\subseteq\mathcal{G}^{-1}(d)$.
Thus, by definition, the action $a$ is realisable on the concrete dynamical model, regardless of the current mode of operation.

\paragraph{Robustifying probability intervals}
We render the transition probability intervals robust against any mode in two steps.
First, we compute the transition probability intervals $\mathcal{P}_z$ for the iMDP $\mathcal{M}^z_\mathbb{I}$ for each individual mode $z \in \mathcal{Z}$.
For each transition $(s,a,s')$, the robust interval $\mathcal{P}_{\forall z} (s,a)(s')$ is then obtained as the \emph{smallest probability interval} that contains the intervals in $\mathcal{P}_z$ for all modes $z \in \mathcal{Z}$:
\begin{equation}
    \mathcal{P}_{\forall z}
    (s,a)(s') = \big[
    \min_{z \in \mathcal{Z}} \underline{p_{z}}(s,a)(s') , \enskip
    \max_{z \in \mathcal{Z}} \overline{p_{z}}(s,a)(s')
    \big] ,
    \label{eq:robust_intervals}
\end{equation}
where $\underline{p_z}(s,a)(s')$ and $\overline{p_{z}}(s,a)(s')$ are again the lower/upper bound probabilities of iMDP $\mathcal{M}^z_\mathbb{I}$.
Using \cref{eq:robust_intervals} we obtain probability intervals that are, by construction, a sound overapproximation of the probability intervals \emph{under any mode $z \in \mathcal{Z}$}.
We use this key observation to state the correctness of the resulting iMDP.

\begin{theorem}[Robust iMDP with unknown mode jumps]
\label{thm:sound}
        The robust iMDP $\mathcal{M}^{\forall z}_\mathbb{I} = (\mathcal{S}, \mathcal{A}_{\forall z}, s_I, \mathcal{P}_{\forall z})$ with actions defined through \cref{eq:all_back} and intervals defined by \cref{eq:robust_intervals} is a $\beta'$-iMDP abstraction for the MJLS in \cref{eq:LinearSystem}, which models state transitions robustly against any mode transition.
        In particular,
        \begin{equation}
            \mathbb{P}^{NV} \Big\{ 
                 (\mathcal{M}^{\forall z}_\mathbb{I} \models_\pi \Phi) \implies
                (\MJLS \models_F \Phi)
            \Big\} \geq 1 - \beta'.
        \end{equation}%
\end{theorem}

We again provide the full proof in \cref{app:proofs}, while only providing an outline here.
The robust iMDP is composed of intervals that contain the true transition probabilities, with a probability of at least $1-\frac{\beta}{|\mathcal{A}|\cdot|\mathcal{S}|}$.
Thus, every probability interval of the robust iMDP contains the intervals for all modes $z \in \mathcal{Z}$ with probability at least $1-\frac{\beta}{|\mathcal{A}|\cdot|\mathcal{S}|} \cdot|\mathcal{Z}|$.
Since the robust iMDP has $|\mathcal{S}| \cdot |\mathcal{A}|$ unique intervals, it follows that all intervals of the iMDP are correct with probability at least
$1-\frac{\beta \cdot |\mathcal{A}| \cdot |\mathcal{S}| \cdot |\mathcal{Z}|}{|\mathcal{A}| \cdot |\mathcal{S}|} = 1-\beta'$.
Analogous to the proof of \cref{thm:controlled_sound}, it is then straightforward to prove that this abstraction is also a $\beta'$-iMDP. 

\section{Synthesis for General PCTL Formulae}
\label{subsec:main:pol}

\begin{figure}[!t]
    \centering
    \begin{minipage}[b]{.48\textwidth}
    \centering
    \resizebox{.88\linewidth}{!}{
    \begin{tikzpicture}[node distance=0.7cm]
        \node (root) {$\Phi$};
        \node (final_and) [below of=root] {$\wedge$};
        \node (P_left) [below of = final_and, xshift=-1.5cm] {$\mathsf{P}_{\geq0.6}$};
        \node (P_r) [below of = final_and, xshift=1.5cm] {$\mathsf{P}_{\geq0.9}$};
        \node (X) [below of = P_left] {$\mathsf{X}$};
        \node (P_l_l) [below of = X] {$\mathsf{P}_{\leq 0.5}$};
        \node (U_left) [below of = P_l_l] {$\mathsf{U}^{\leq K-1}$};
        \node (neg_left) [below of = U_left, xshift=-1cm] {$\neg$};
        \node (or_left) [below of = U_left, xshift = 1cm] {$\vee$};
        \node (C_left_left) [below of = neg_left] {$T_C$};
        \node (L_left) [below of = or_left, xshift = -0.5cm] {$T_L$};
        \node (C_left_right) [below of = or_left, xshift=0.5cm] {$T_C$};
        
        \node (U_right) [below of = P_r] {$\mathsf{U}^{\leq K}$};
        \node (neg_right) [below of = U_right, xshift=-1cm] {$\neg$};
        \node (C_right) [below of = neg_right] {$T_C$};
        \node (G) [below of = U_right, xshift=1cm] {$T_G$};
    
        \draw [-] (root) -- (final_and);
        \draw [-] (final_and) -- (P_left);
        \draw [-] (final_and) -- (P_r);
        \draw [-] (X) -- (P_left);
        \draw [-] (X) -- (P_l_l);
        \draw [-] (U_left) -- (P_l_l);
        \draw [-] (U_left) -- (neg_left);
        \draw [-] (C_left_left) -- (neg_left);
        \draw [-] (U_left) -- (or_left);
        \draw [-] (C_left_right) -- (or_left);
        \draw [-] (L_left) -- (or_left);
    
        \draw [-] (U_right) -- (P_r);
        \draw [-] (U_right) -- (neg_right);
        \draw [-] (U_right) -- (G);
        \draw [-] (C_right) -- (neg_right);
    \end{tikzpicture}
    }
    \caption{Parse tree for PCTL formula \cref{eq:pctl_tree}.}
    \label{fig:parse_tree}

    \end{minipage}%
    \hfill
    \begin{minipage}[b]{0.48\textwidth}
        \centering
        \includegraphics[width=.95\linewidth]{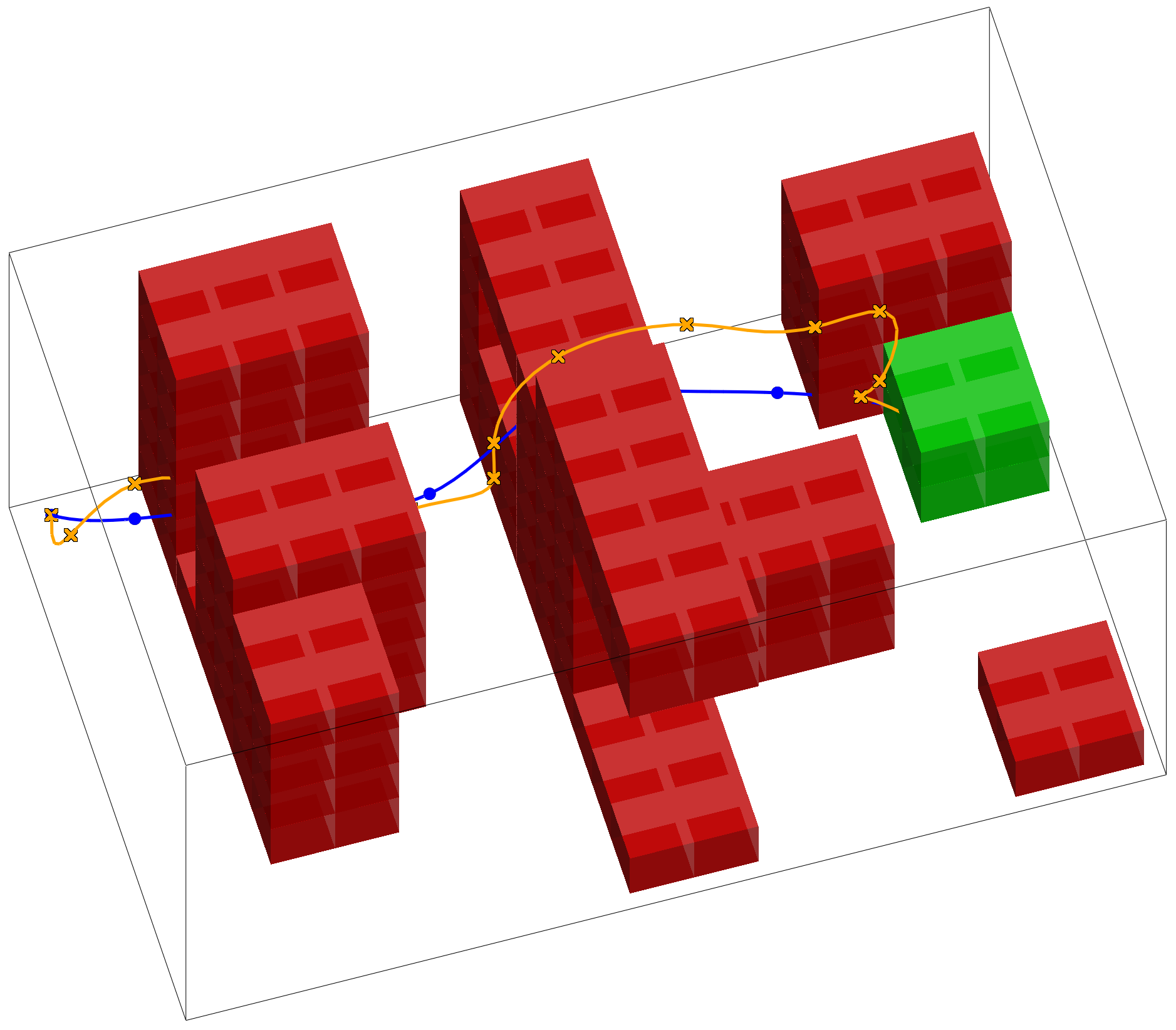}
        \caption{Simulated paths under weak (blue) and strong (orange) wind for the drone.}
        \label{fig:drone} 
    \end{minipage}
\end{figure}

To synthesise optimal policies in our discrete abstraction, we use the probabilistic model checker PRISM~\cite{kwiatkowskaPRISMVerificationProbabilistic2011}.
We handle complex and nested PCTL formulae by defining a \emph{parse tree}~\cite{baierPrinciplesModelChecking2008a}, whose leaves are atomic propositions, and whose branches are logical, temporal, or probabilistic operators. 
The complete formula can be verified using a bottom-up approach. 
As an example, consider the formula
\begin{equation}
    \Phi = \mathsf{P}_{\geq0.6}[\mathsf{X}\mathsf{P}_{\leq 0.5}(\neg T_C \mathsf{U}^{\leq K-1}(T_L \vee T_C))] \wedge \mathsf{P}_{\geq 0.9}[\neg T_C \mathsf{U}^{\leq K} T_G],
    \label{eq:pctl_tree}
\end{equation}
with atomic propositions $T_C, T_L, T_G$.
The parse tree for this formula is shown in \cref{fig:parse_tree}.
We will explain and use this formula in the temperature control experiment in \cref{sec:results}.
When considering multiple PCTL fragments, we find a policy associated with each fragment (our example above will find two policies, one satisfying $\mathsf{P}_{\geq0.6}[\mathsf{X}\mathsf{P}_{\leq 0.5}(\neg T_C \mathsf{U}^{\leq K-1}(T_L \vee T_C))]$ the other $\mathsf{P}_{\geq 0.9}[\neg T_C \mathsf{U}^{\leq K} T_G]$).
At runtime, we choose which PCTL fragment to satisfy and apply its associated policy as $\pi^\star$. 

\subsection{Unsatisfied Formulae}
If the PCTL formula is not satisfied by the iMDP, we refine our abstraction by increasing the number of samples used to compute the probability intervals (shown by the dashed line in \cref{fig:overview}).
As also discussed in more detail and shown experimentally by \cite{badingsSamplingBasedRobustControl2021}, this refinement tightens the probability intervals, which in turn improves the probability of satisfying the property.
We iteratively refine our abstraction until the formula is satisfied or until a maximum number of iterations is exceeded (which we fix a priori), in which case nothing is returned.
In this way, our method is sound, but not complete: if the formula is not satisfied after the maximum number of iterations, this in general does not imply that the formula cannot be satisfied at all.
However, for any policy that is returned by our algorithm, the correctness result of \cref{thm:sound} holds.

\subsection{Controller Synthesis via Policy Refinement}
\label{subsec:main:controller}

We refine the optimal policy $\pi^\star$ to obtain a hybrid-state feedback controller $F$ for the MJLS, as follows.
Given the current continuous state $x \in \mathcal{X}$, mode $z \in \mathcal{Z}$ and time step $k \in \mathbb{N}$, we first find the element $X_i$ of partition $X$ containing state $x$, such that $x \in X_i$.
Depending on whether we consider abiding by modelling \cref{ass:MDP} or \cref{ass:unknown}, we then proceed as follows:
\begin{itemize}
    \item For \cref{ass:MDP}, we find the product state $s_\times = (z,s)$ associated with the current mode $z \in \mathcal{Z}$ and state $s$.
    We then look up the optimal product action $a_\times = \pi^\star(s_\times, k) = (b,a)$ from policy $\pi^\star$, with corresponding switching action $b$ and continuous action $a$.
    \item For \cref{ass:unknown}, it suffices to know state $s$ associated with $X_i$ only, and we directly obtain action $a = \pi^\star(s, k)$, with no switching action.
\end{itemize}%
Finally, we compute the continuous control input $u$ associated with action $a$ by calculating the control input that drives us to the associated target point $d$, using $u = B_z^+(d-A_zx-q_z)$, with $B_z^+$ representing the pseudoinverse of $B_z$.
\section{Numerical Experiments}
\label{sec:results}

We have implemented our techniques in Python, using the probabilistic model checker PRISM \cite{kwiatkowskaPRISMVerificationProbabilistic2011} to verify the satisfaction of PCTL formulae on iMDPs.
The codebase is available at \href{https://github.com/lukearcus/ScenarioAbstraction}{https://github.com/lukearcus/ScenarioAbstraction}.
Experiments were run on a computer with 6 3.7 GHz cores and 32 GB of RAM.
We demonstrate our techniques on two models: (1) a UAV motion control problem with two possible levels of noise, and (2) a building temperature regulation problem, in line with our running example from \cref{sec:background}. 
Details on the UAV model and additional experimental results can be found in \cref{app:add_exp}.

\subsection{UAV Motion Planning}
We consider a more refined, \emph{hybrid} version of the unmanned aerial vehicle (UAV) motion planning problem from \cite{badingsSamplingBasedRobustControl2021}. 
We consider two discrete modes, which reflect different levels of noise, namely low and high wind speeds.
We use our framework considering \cref{ass:MDP}.
The PCTL specification $\Phi = \mathsf{P}_{\geq0.5}[\neg O \mathsf{U}^{\leq K} G]$ requires reaching a goal set $G$ (highlighted in green in \cref{fig:drone}), whilst avoiding obstacles $O$ (highlighted in red).
We choose a finite time horizon $K = 64$.
While our theoretical contributions hold for any probability distribution for the additive noise, in this particular experiment we sample from a Gaussian.

\paragraph*{Scalability}
The number of iMDP states equals the number of partitions, multiplied by the number of discrete modes, here resulting in 51,030 states.
The number of transitions depends on the number of samples: with 100 samples, we generate an iMDP with 92.7 million transitions; with 200 samples, 154 million transitions. 
Computing the iMDP actions enabled  in the abstraction is independent of sampling and takes 8.5 min; 
computing the transition probability intervals of the iMDP takes 70 min; 
formal synthesis of the optimal policy takes 40 s, and control refinement occurs online. 

\paragraph*{Variable noise affects decisions}
With our techniques, we synthesise a controller that accounts for different noise levels at runtime and reasons about the probability of the noise level changing.
Thus, our framework makes use of the information available regarding the jump process, while at the same time reasoning explicitly over the stochastic noise affecting the continuous dynamics in each mode.

\subsection{Temperature Regulation in a Building}
We consider again the 2-room building temperature control problem \cite{abateProbabilisticReachabilitySafety2008} introduced in \cref{example:temp1}.
Recall that the state $x = [T_1, T_2]^\top \in \mathbb{R}^2$ models the temperature in both rooms, and the control input (modelling the power supply to the heaters) is constrained to $u \in \mathbb{R}^2$.
The values of the constants in \cref{eq:example:dynamics} are $a_{12} = 0.022, \ b_1 = b_2 = 0.0167, \ k_f = 0.8, \ k_r = 0.4, \ x_a = 6$.
The noise is distributed according to a zero-mean Gaussian with a standard deviation of 0.2.

\begin{figure}[!t]
    \begin{minipage}[b]{.3\textwidth}
        \centering
        \includegraphics[width=\linewidth]{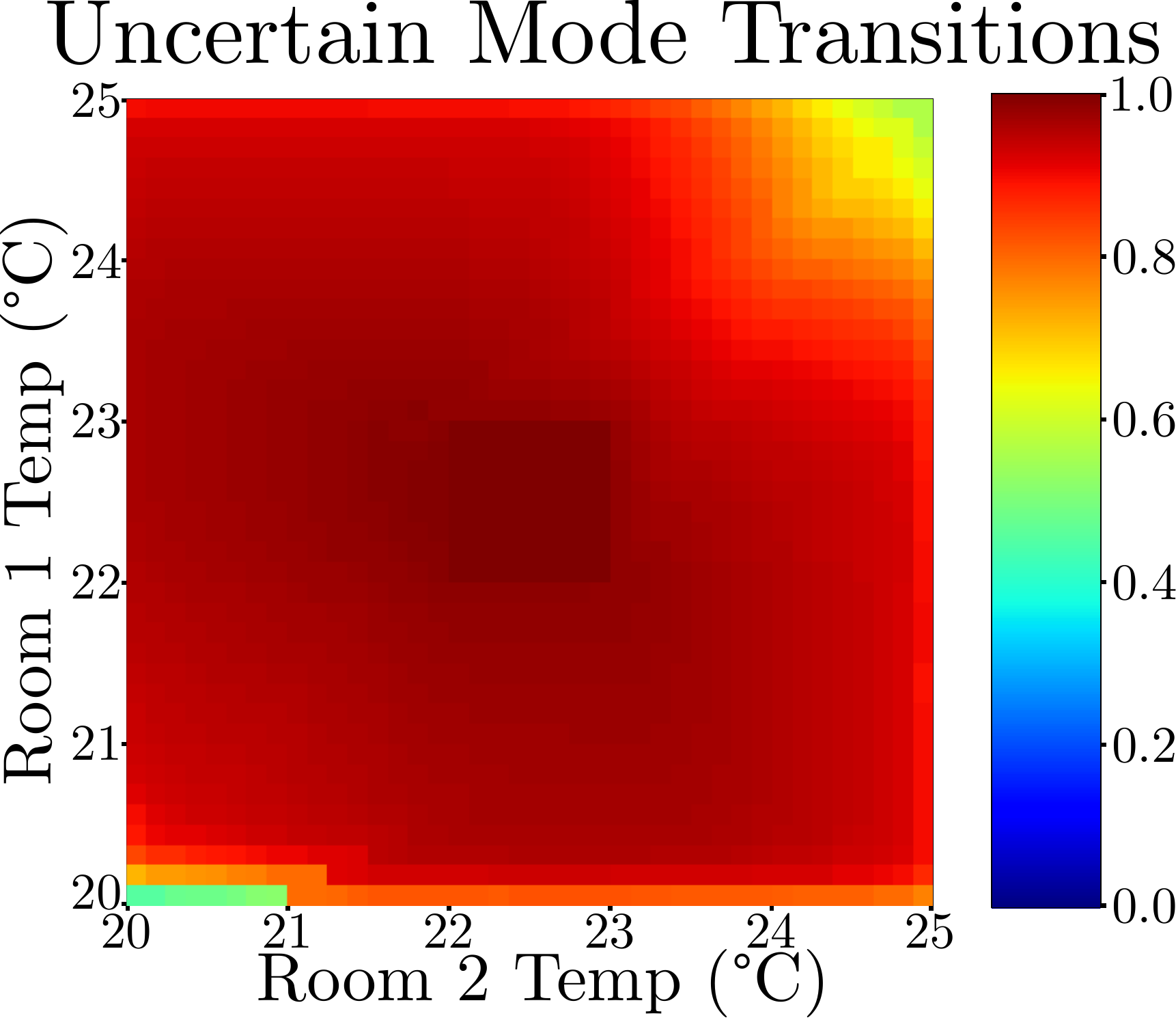}
        \caption{Lower bound satisfaction probabilities with interval mode jump probabilities (\cref{ass:MDP}).}
        \label{fig:heat_mdp}
    \end{minipage}
    \hfill
    \begin{minipage}[b]{.3\textwidth}
        \centering
        \includegraphics[width=\linewidth]{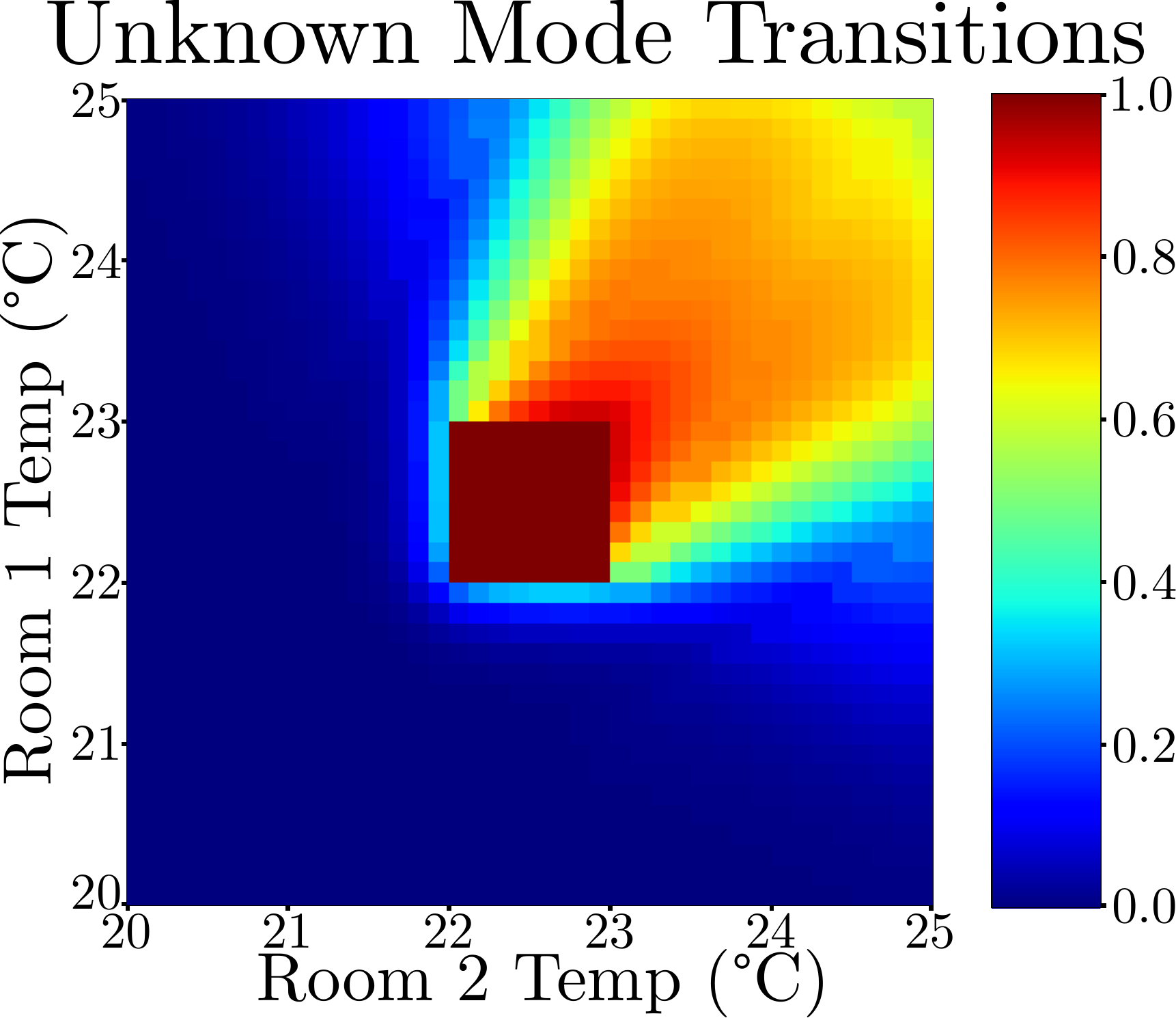}
        \caption{Lower bound satisfaction probabilities with unknown mode jump probabilities (\cref{ass:unknown}).}
        \label{fig:heat_robust}
    \end{minipage}
    \hfill
    \begin{minipage}[b]{.26
    \textwidth}
        \centering
        \includegraphics[width=\linewidth]{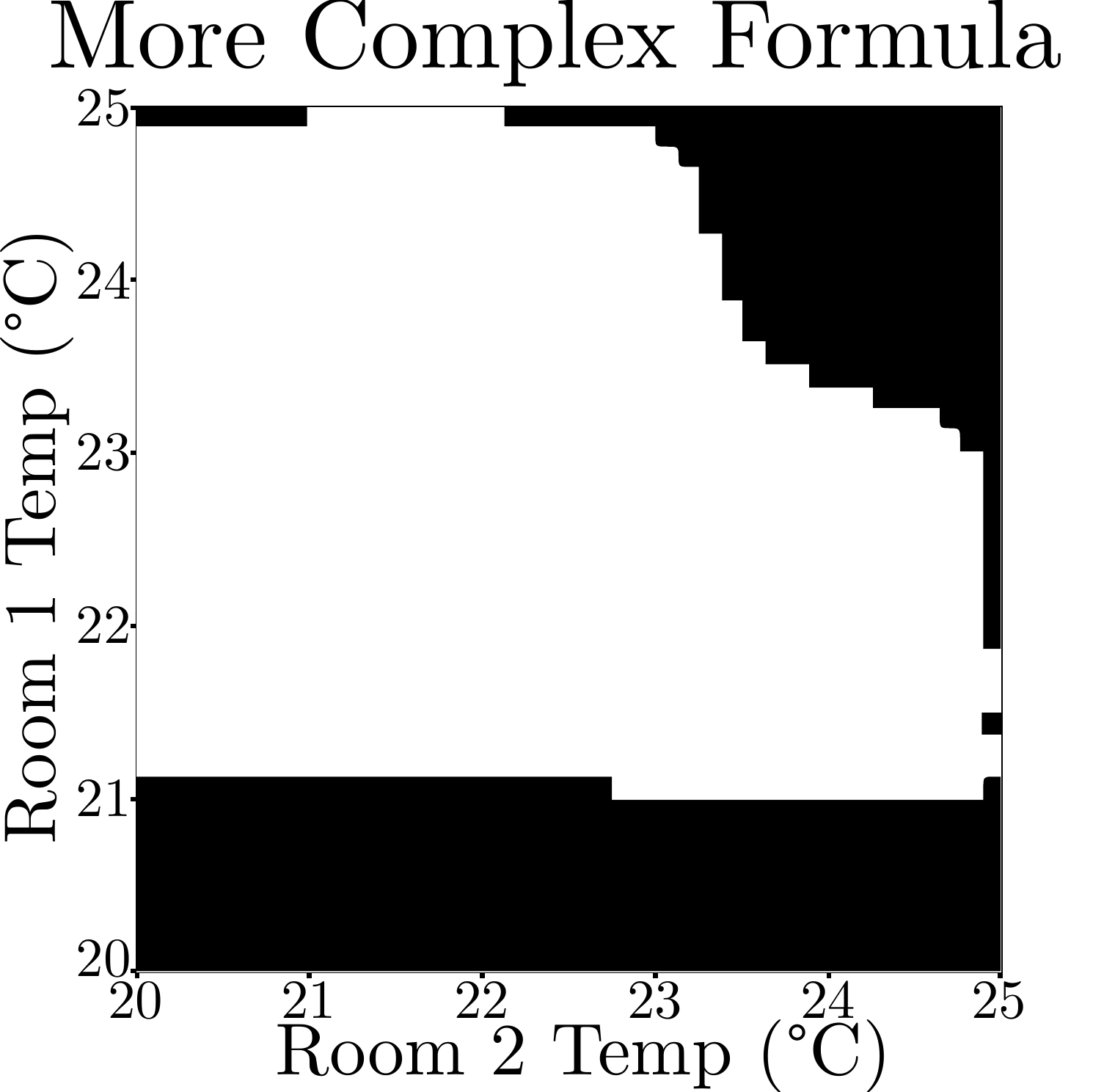}
        \caption{States that satisfy (in white) the general PCTL formula in \cref{eq:pctl_tree} (\cref{ass:MDP}).}
        \label{fig:complex}
    \end{minipage}
    \vspace{-0.5cm}
\end{figure}

We wish to optimise the probability of satisfying the path formula $\psi = (\neg T_C)\mathsf{U}^{\leq K}(T_G)$, with goal temperature $T_G$ between 22 and 23$^\circ$C, and critical temperature $T_C$ less than 20$^\circ$C or greater than 25$^\circ$C.
We partition the state space into 1600 regions, using a time horizon $K=32$. 
We look into two setups, one fulfilling \cref{ass:MDP} (\cref{fig:heat_mdp}) and the other \cref{ass:unknown} (\cref{fig:heat_robust}).
We show the results for all initial continuous states, and in \cref{fig:heat_mdp} we consider starting in mode 1 (whereas the bounds in \cref{fig:heat_robust} hold for any initial mode).

\paragraph*{Assumptions affect conservativism and scalability}
When we wish to be robust to all possible modes (cf. \cref{ass:unknown}), our generated iMDP is much smaller (with about 18 times fewer transitions), since we have a single robust iMDP, compared to a product iMDP. 
However, as expected and seen in \cref{fig:heat_robust}, the obtained probability lower bounds are much more conservative. 
Thus, compared to \cref{ass:unknown}, \cref{ass:MDP} reduces the level of conservatism (because we exploit the probability intervals of the Markov jump process) at the cost of increasing the size of the abstraction.  

\subsection{Controller Synthesis for General PCTL Formulae} 
We now consider the general PCTL formula in \cref{eq:pctl_tree} to show the applicability of our techniques beyond reach-avoid specifications.
This formula requires (1) heating both rooms to a goal temperature while avoiding critical temperatures; and (2) reaching a state at the next time step, which is able to avoid entering an unwanted or critical temperature. 
The new atomic proposition $T_L$ specifies that temperatures should be kept below 21$^\circ$C in room 1. 
In \cref{fig:complex}, we show the set of iMDP states that satisfy the PCTL formula (shown in white), if the fan heater is initially in room 1 (see \cref{example:temp1}).
Thus, we can compute a feedback controller for the MJLS satisfying the PCTL formula, unless the initial room temperature is (approximately) below $21^\circ$C, or if both initial temperatures are too high.
\section{Conclusions and Future Work}
\label{sec:conclusion}

We have presented a new method for synthesizing certifiably correct controllers for MJLSs with hybrid, stochastic and partly unknown dynamics. 
We considered both the case where an estimate of the switching probabilities across discrete operation modes is known, and the alternative instance where these probabilities are not known at all.
Our experiments have demonstrated the efficacy of our methods on a number of realistic problems.

Future research directions include considering state-dependent mode switches (e.g. for models in \cite{lunzeHandbookHybridSystems2009,AKLP10}), estimating mode-switching probabilities with the scenario approach, and dealing with a setting where matrices are only known to belong to a convex polytope, as in~\cite{Badings2022ProbabilitiesEnough_AAAI} for non-hybrid systems.

\newpage
\appendix
\renewcommand{\thesection}{\appendixname~\arabic{section}}
\section{Proofs}
\label{app:proofs}

\subsection{Proof of \cref{thm:controlled_sound}}

Since the dynamics for each mode $z \in \mathcal{Z}$ is a linear system and mode jumps happen after state transitions, then the one-step backward reachable set is simply the one-step backward reachable of the linear dynamics in the mode.
As such, to define the sets of enabled actions, we can simply use the individual iMDP abstractions for linear dynamics.
By constructing a Cartesian product state and action space, we encompass our state and action in both the current mode iMDP and in the mode switching iMDP.

From the definition of the mode switching iMDP and the construction of the individual mode $\beta$-iMDPs from \cref{sec:AAAI}, we have 
\begin{equation}
\label{eq:proof:thm2:1}
\begin{aligned}
    &T^\star(z,b)(z') \in \mathcal{T}(z,b)(z') = [\underline{t}(z,b)(z'), \overline{t}(z,b)(z')],\\
        &    \mathbb{P}_z^V \Big\{ 
        \truePz(s,a)(s') \in  \mathcal{P}_z(s,a)(s'), \enskip \forall s \in \mathcal{S}
    \Big\} \geq 1 - \frac{\beta}{|\mathcal{S}|\cdot|\mathcal{A}|},
\end{aligned}%
\end{equation}%
with $\mathcal{T}(z,b)(z')$ the probability interval for transition $(z,b,z')$ in the mode switching iMDP $\mathcal{M}_\mathbb{I}$, and with $\mathcal{P}_z(s,a)(s') = [\underline{p_z}(s,a)(s'), \overline{p_z}(s,a)(s')]$ the probability interval for transition $(s,a,s')$ in individual-mode iMDP $\mathcal{M}_\mathbb{I}^z$.
Moreover, recall from \cref{sec:AAAI} that $T^\star$ is the true transition function of the mode switching MDP, and that $\truePz$ is the true transition function of the MDP abstraction for mode $z \in \mathcal{Z}$.

Based on the definition of the transition function $\mathcal{P}_\times$ of the product iMDP in \cref{eq:product_imdp_interval}, it immediately follows that for each $(z,s), (z', s') \in \mathcal{Z} \times \mathcal{S}$ and $(b,a) \in \mathcal{A}(z,s)$, it holds that
\begin{equation}
\label{eq:proof:thm2:2}
\begin{aligned}
&\mathbb{P}^{NV}\{\truePtimes((z,s)(b,a))((z',s')) \in \mathcal{P}_\times((z,s)(b,a))((z',s'))\} \geq 1-\frac{\beta}{|\mathcal{S}|\cdot|\mathcal{A}|},
\end{aligned}
\end{equation}
where $\truePtimes$ is the true (unknown) transition function of the product MDP.
Since the iMDP can be shown to have at most $|\mathcal{S}|\cdot|\mathcal{A}|\cdot |\mathcal{Z}|$ unique transition probability intervals, we can make use of Boole's inequality to show that with probability at least $1-\beta'$, all probabilities are contained within their intervals:
\begin{equation}
\label{eq:proof:thm2:3}
\begin{aligned}
        \mathbb{P}^{NV} \Big\{ & \truePtimes((z,s)(b,a))((z',s')) \in \mathcal{P}_\times((z,s)(b,a))((z',s')), \\
        & \forall (z,s),(z',s') \in \mathcal{Z} \times \mathcal{S}, \, \forall (b,a) \in \mathcal{A}(z,s) \Big\} \geq 1 - \frac{\beta\cdot|\mathcal{S}|\cdot|\mathcal{A}|\cdot|\mathcal{Z}|}{|\mathcal{S}|\cdot|\mathcal{A}|}=1-\beta',
\end{aligned}%
\end{equation}%
which we write as $\mathbb{P}^{NV}\{\mathcal{M}^{\truePtimes} \in \mathcal{M}_\mathbb{I}^\times \} \geq 1-\beta'$ for brevity.

In order to connect this result on the abstract system back to the dynamical system, we consider again the mapping from hybrid states to abstract states, and the MDP $\mathcal{M}_\times^{\truePtimes}$ with transition probabilities $\truePtimes((z,s)(b,a))((z',s'))$ which are the true transition probabilities.
We now demonstrate that this mapping from hybrid states to abstract states induces a probabilistic feedback refinement~\cite{hermannsProbabilisticLogicalCharacterization2011}.
A sufficient condition for this relation is that, for any pair of related states $((z(k),x(k)), s_\times)$, and for all actions $a_\times \in \mathcal{A}_\times(s_\times)$ enabled in that joint state $s_\times$, there exist inputs $b \in \mathcal{B}, u \in \mathcal{U}$ such that the probability of transitioning to any state $s_\times'$ in the MDP is equal to the probability of transitioning to any $x' \in X_i$ and to $z'$ (where $X_i$ is the element of the partition associated with $s_\times'$).
First, by definition, it is evident that the initial states match so that $(z(0),x(0))$ maps to $s^I_\times$.
Then, for any pair of states $((z(k),x(k)), s_\times)$, we have that
\begin{align}
\label{eq:proof:thm2:4}
        &\forall a_\times \in \mathcal{A_\times}, s_\times' \in \mathcal{S_\times} \colon
        \nonumber
        \\
        &\truePtimes(s_\times,a_\times)(s_\times') = \truePz(s,a)(s') \cdot T^\star(z,b)(z') =
        \\
        &\left[\int_{\mathbb{R}^n}\mathbbm{1}_{X_i}\left(A_{z(k)}x(k)+B_{z(k)}F(x(k),z(k),k)+q_{z(k)}+\xi\right) \mathbb{P}_{w_{z(k)}}(d\xi)\right]\cdot T^\star(z,b)(z')
        \nonumber
\end{align}%
where $\truePz$ is the true transition probability function for mode $z$, $T^\star$ is the true mode switch probability function, $\mathbbm{1}_{X_i}(\cdot)$ is the indicator function (which returns value 1 if, and only if, its argument belongs to the set $X_i$), and $X_i$ is the element of the partition associated with $s_\times'$ (and hence also the element of the partition associated with $s'$).
Importantly, this relation implies that any PCTL formula satisfiable for the MDP is also satisfiable for the dynamical system~\cite{hermannsProbabilisticLogicalCharacterization2011}.
More formally, for this product MDP $\mathcal{M}^{\truePtimes}$ with transition function $\truePtimes$, it holds that
\begin{equation}
    \label{eq:proof:thm2:5}
    (\mathcal{M}^{\truePtimes} \models_\pi \Phi) \implies (\MJLS \models_F \Phi).
\end{equation}
Finally, combining \cref{eq:proof:thm2:3,eq:proof:thm2:5}, we arrive at the desired expression, namely
\begin{equation}
    \label{eq:proof:thm2:6}
     \mathbb{P}^{NV}\{ (\mathcal{M}_\mathbb{I}^\times \models_\pi \Phi) \implies (\MJLS \models_F \Phi)\} \geq 1-\beta'.
\end{equation}
Thus, the generated abstraction is a $\beta'$-iMDP.

\subsection{Proof of \cref{thm:sound}}

Consider the robust iMDP $\mathcal{M}_\mathbb{I}^{\forall z} = (\mathcal{S}, \mathcal{A}_{\forall z}, s_I, \mathcal{P}_{\forall z})$ obtained via the procedure outlined in \cref{subsec:main:AssumptionB}.
Since we define the backward reachable set as the intersection of the individual backward reachable sets, then a partition $X_i \subseteq \mathcal{G}^{-1}(d) \subseteq \mathcal{R}^{-1}_z(d), \forall z \in \mathcal{Z}$ and for all target points $d$.

On the concrete model at every time step, we can measure the current mode, thus we can calculate the control input to drive our noiseless successor state to $d$ as $u = B_z^+(d-A_zx-q_z)$, with $B_z^+$ representing the pseudoinverse of B (as also described in \cref{subsec:main:controller}).
By the definition of the backward reachable set, this control input will be a valid control input such that $u \in \mathcal{U}$, regardless of the mode.

We choose each transition probability interval as the smallest interval containing \emph{all} individual mode probability intervals, so $\mathcal{P}_z(s,a)(s') \subseteq \mathcal{P}_{\forall z}(s,a)(s'), \forall z \in \mathcal{Z}$.
Each individual interval is generated with the scenario approach.
Thus, for the true transition probability $\truePz(s,a)(s')$, we have for all $z \in \mathcal{Z}$ that
\begin{equation}
    \label{eq:proof:thm3:1}
    \mathbb{P}^V\{\truePz(s,a)(s') \in \mathcal{P}_z(s,a)(s'), 
    \enskip \forall s \in \mathcal{S} \} \geq 1-\frac{\beta}{|\mathcal{S}|\cdot|\mathcal{A}|}.
\end{equation}
It follows that, since state $s$ is contained in the backward reachable set of $d$ in mode $z$, and $\mathcal{P}_z(s,a)(s') \subseteq \mathcal{P}_{\forall z}(s,a)(s')$ then $\mathbb{P}^V\{\truePz(s,a)(s') \in \mathcal{P}_{\forall z}(s,a)(s'), \forall s \in \mathcal{S}\} \geq 1-\frac{\beta}{|\mathcal{S}|\cdot|\mathcal{A}|}$.
Since we wish to be robust to all possible modes, we have that
\begin{equation}
    \label{eq:proof:thm3:2}
    \mathbb{P}^{NV}\{\truePz(s,a)(s') \in \mathcal{P}_{\forall z}(s,a)(s'), \enskip
    \forall s \in \mathcal{S}, \forall z \in \mathcal{Z}\} \geq 1-\frac{\beta\cdot|\mathcal{Z}|}{|\mathcal{S}|\cdot|\mathcal{A}|}.
\end{equation}
Lastly, by observing that the iMDP has at most $|\mathcal{S}|\cdot|\mathcal{A}|$ unique probability intervals, we can again use Boole's inequality to show that with probability at least $1-\beta'$, all probabilities are contained within their intervals
\begin{equation}
\begin{aligned}
        \mathbb{P}^{NV} \Big\{ \truePz(s,a)(s') & \in \mathcal{P}_{\forall z}(s,a)(s'),
        \forall s,s' \in \mathcal{S}, a \in \mathcal{A}, z \in \mathcal{Z} \Big\} \\ 
        & \geq 1 - \frac{\beta\cdot|\mathcal{S}|\cdot|\mathcal{A}|\cdot|\mathcal{Z}|}{|\mathcal{S}|\cdot|\mathcal{A}|}=1-\beta'.
\end{aligned}
\end{equation}

As noted above, on the concrete system, we are always able to calculate a valid control input. 
Thus, for any possible evolution of the mode iMDP and at every time step, we can calculate a valid control input, and we know that with a confidence of $1-\beta'$, the probability of transitioning to $s'$ is in the interval $\mathcal{P}_{\forall z}(s,a)(s')$, again regardless of which mode we do uncover.
We can thus conclude, for any possible mode-switching MDP, our resulting abstraction will properly contain the true transition kernel of the continuous system.

We again consider the concepts of probabilistic feedback refinements~\cite{hermannsProbabilisticLogicalCharacterization2011}.
Our mapping now maps the continuous part $x$ of the hybrid state in the MJLS to a single state $s$, regardless of mode $z$; thus we consider a pair of states $(x,s)$.
For all such pairs of states, and for every mode, we then have
\begin{equation}
\begin{aligned}
        & \forall a\in \mathcal{A}, s' \in \mathcal{S}, z \in \mathcal{Z}\colon\\
        & \qquad \truePz(s,a)(s') = 
       \int_{\mathbb{R}^n}\mathbbm{1}_{X_i}\left(A_{z}x(k)+B_{z}F(x(k),z,k)+q_{z}+\xi\right) \mathbb{P}_{w_{z}}(d\xi),
\end{aligned}
\end{equation}
where $X_i$ is the element of the partition associated with $s'$.

Consider now an iMDP $\mathcal{M}_\mathbb{I}^{\truePall}$, with intervals $$\mathcal{P}^\star_{\forall z}(s,a)(s') = [ \min_{z \in \mathcal{Z}}\truePz(s,a)(s'), \max_{z \in \mathcal{Z}}\truePz(s,a)(s')],$$ such that $\truePz(s,a)(s') \in \mathcal{P}^\star_{\forall z}(s,a)(s'), \forall z \in \mathcal{Z}$, states and actions are identical to those for $\mathcal{M}_\mathbb{I}^{\forall z}$.
Since this true robust iMDP contains the true transition probabilities $\truePz$ of every mode, then satisfaction of a formula on this iMDP must imply satisfaction of the formula on any concrete model with the known mode dynamics, but any mode-switching MDP, so that
$$(\mathcal{M}_\mathbb{I}^{\truePall} \models_\pi \Phi) \implies (\MJLS \models_F \Phi).$$
Then, this iMDP is contained within our generated iMDP with a confidence $1-\beta'$
$$\mathbb{P}^{NV} \Big\{\mathcal{M}_\mathbb{I}^{\truePall} \in \mathcal{M}_\mathbb{I}^{\forall z} \Big\} \geq 1-\beta',$$
so that we can finally conclude
\begin{equation}
     \mathbb{P}^{NV}\{ (\mathcal{M}_\mathbb{I}^{\forall z} \models_\pi \Phi) \implies (\MJLS \models_F \Phi)\} \geq 1-\beta'.
\end{equation}
Thus, the generated abstraction is a $\beta'$-iMDP.
\section{Experiment Details and Additional Results}
\label{app:add_exp}

In this appendix, we first provide the explicit model formulation for the UAV experiments.
Thereafter, we present additional results for the temperature control benchmark in particular.

\subsection{UAV motion planning}

For the UAV motion planning problem, we considered a 6-dimensional state vector defined as $x = (p_x, p_y, p_z, v_x, v_y, v_z)^\top \in \mathbb{R}^6$, with $p_i$ and $v_i$ denoting the position and velocity in direction $i$. 
The dynamics follow a double integrator model, so that the resulting state equations are:
\begin{equation}
\label{eq:UAV_model}
    x(k+1) =
    \begin{bmatrix}
        1 & 0 & 0 & T & 0 & 0 \\
        0 & 1 & 0 & 0 & T & 0 \\
        0 & 0 & 1 & 0 & 0 & T \\
        0 & 0 & 0 & 1 & 0 & 0 \\
        0 & 0 & 0 & 0 & 1 & 0 \\
        0 & 0 & 0 & 0 & 0 & 1
    \end{bmatrix}
    x(k) + 
    \begin{bmatrix}
        \frac{T^2}{2} & 0 & 0 \\
        0 & \frac{T^2}{2} & 0 \\
        0 & 0 & \frac{T^2}{2} \\
        T & 0 & 0 \\
        0 & T & 0 \\
        0 & 0 & T
    \end{bmatrix}
    u(k) + w(k),
\end{equation}
where $w(k)$ is the noise term arising from turbulence, $T$ is the discretization constant, and $u(k) \in \mathbb{R}^3$ is the acceleration, and control input, which is constrained to the interval $[-4,4]$.

Since the model in \cref{eq:UAV_model} is not fully actuated (it has only 3 control inputs, with a state space of 6 dimensions) we group every two time steps together to rewrite the model as
\begin{equation}
    x(k+2) = \bar{A}x(k) + \bar{B}\begin{bmatrix}u(k)\\u(k+1)\end{bmatrix} + \begin{bmatrix}Aw(k)\\w(k+1)\end{bmatrix}.
\end{equation}
With $\bar{A} = A^2$ and $\bar{B} = \begin{bmatrix} AB&B\end{bmatrix}$.
Then, the two control inputs are placed into a single vector, which now has dimension 6.

In this setup, the only difference between the two modes is the distribution of the noise $w(k)$: in the low wind speed mode, this is distributed according to a zero-mean Gaussian with standard deviation of 0.15. 
In the high wind speed mode, the standard deviation is instead 1.5.
The jumps between these modes are modelled with a precisely known MDP, with a 10\% chance of switching from low to high wind speed, and then a 30\% chance of switching from high to low.

The objective, is to reach the goal region (highlighted in green in \cref{fig:drone}), whilst avoiding critical regions (highlighted in red in \cref{fig:drone}), within 64 time steps (or 32 time steps with the amended model).

\subsection{Additional Outcomes for Temperature Control}

\begin{figure}[b!]
    \centering
    \includegraphics[width=0.6\textwidth]{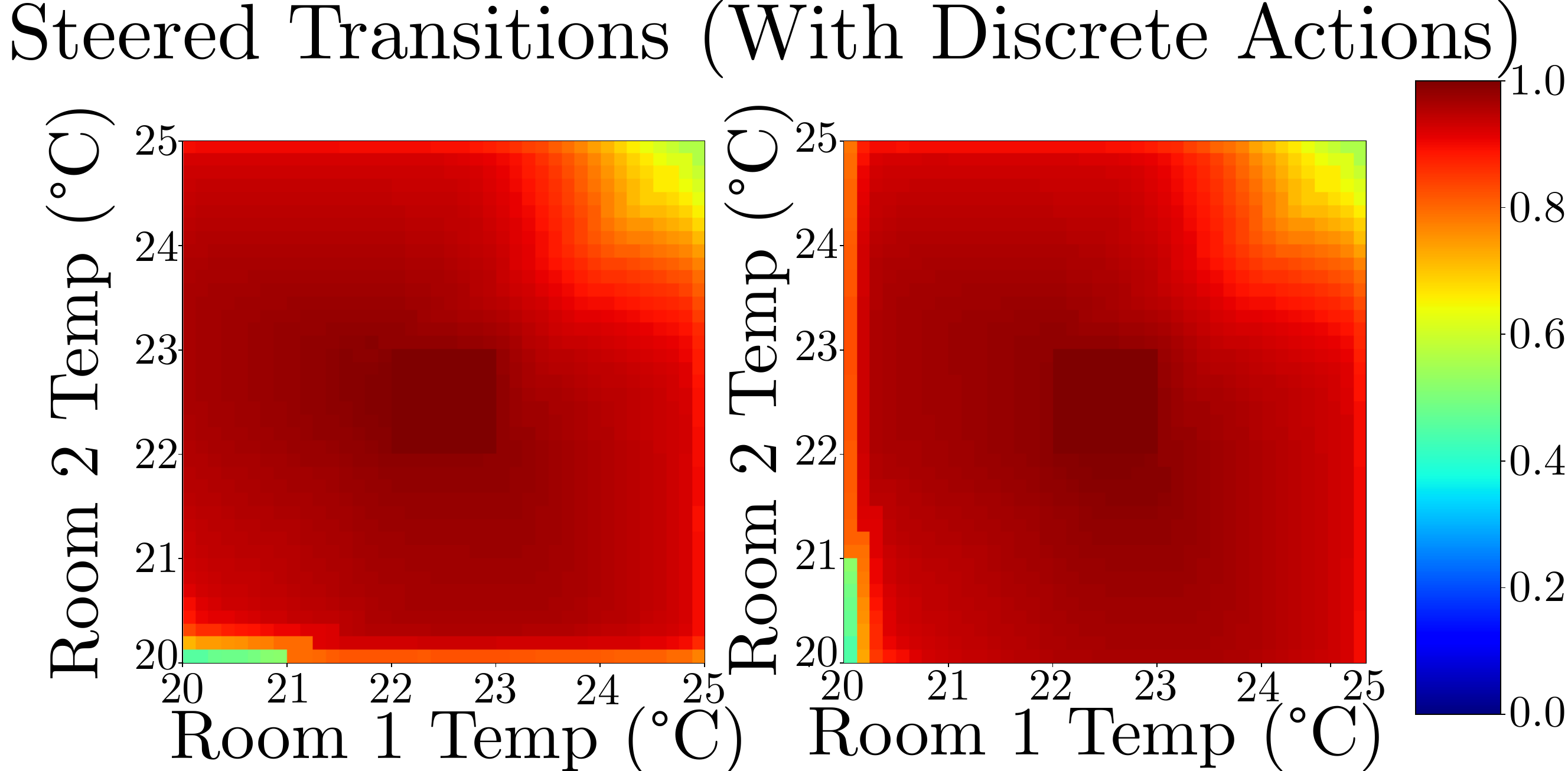}
    \caption{Lower bound probabilities on reaching the goal temperature, as in \cref{fig:heat_mdp}, but now for both initial modes for the MJLS (left: initial mode 1; right: initial mode 2).}
    \label{fig:heat_mdp_all}
\end{figure}

\begin{figure}[b!]
    \centering
    \includegraphics[width=0.6\textwidth]{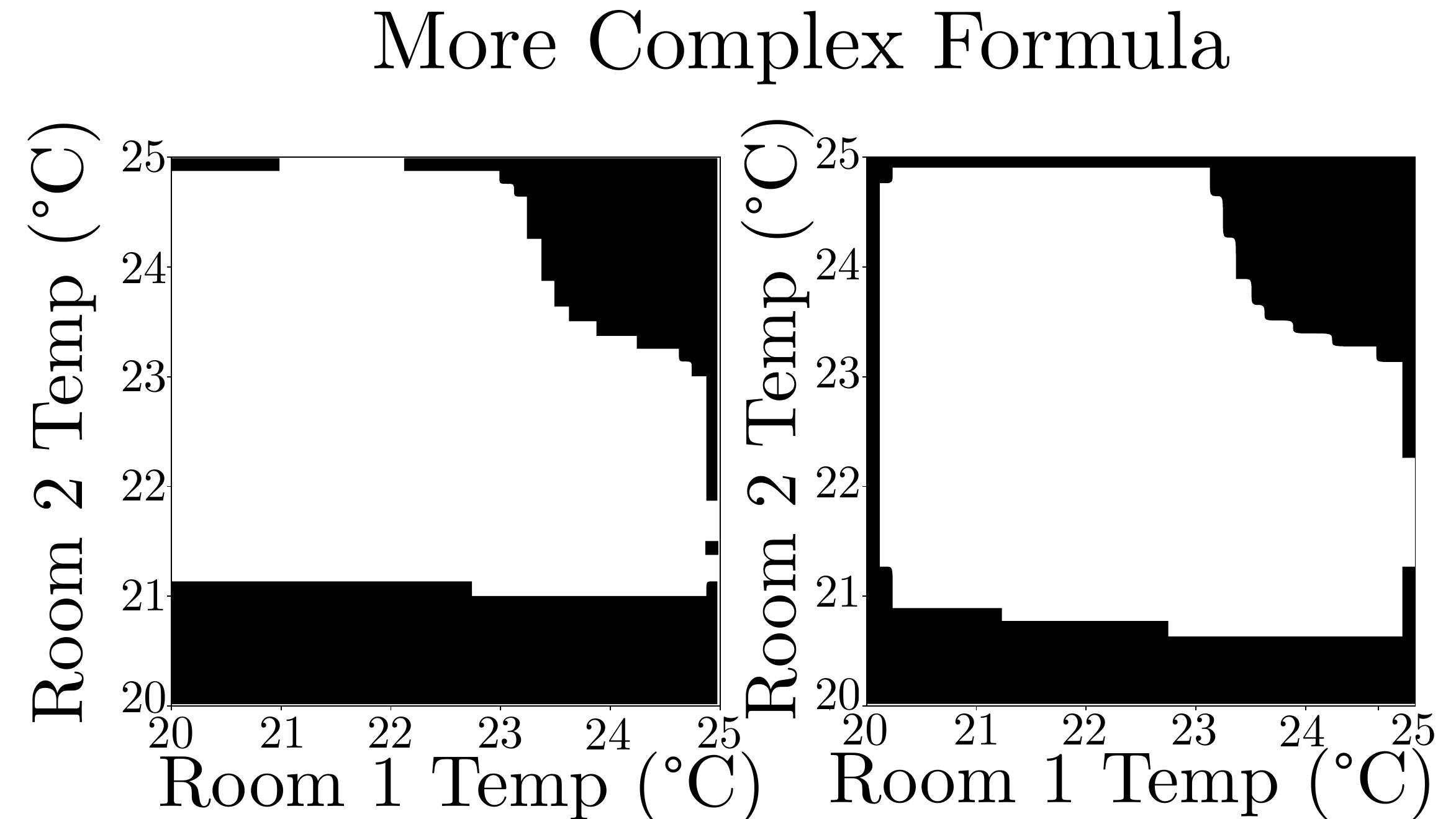}
    \caption{Initial states that satisfy the PCTL formula in \cref{eq:pctl_tree} (shown in white), presented for both initial modes of the MJLS (left: initial mode 1; right: initial mode 2).}
    \label{fig:complex_all}
\end{figure}

\subsubsection{Initial mode affects obtained controllers} 
In \cref{fig:heat_mdp_all,fig:complex_all}, we present the results for the building temperature control problem, if the initial MJLS mode is either 1 or 2.
From \cref{fig:heat_mdp_all}, we observe that the initial mode indeed affects the probability of reaching the satisfying the PCTL property.
The clearest evidence for this is when initial temperatures are low in both rooms.
The disparity arises from the fact that we have less control over the other room, so when starting in mode 1, we are more likely to fail to meet the specification if mode 2 has a close to critical temperature, and vice versa for starting in mode 2.
This demonstrates that finding the true initial mode is important for the generated guarantees, analysing the results for an incorrect initial mode will lead to incorrect conclusions.

\begin{figure}[t!]
    \centering
    \includegraphics[width=0.6\textwidth]{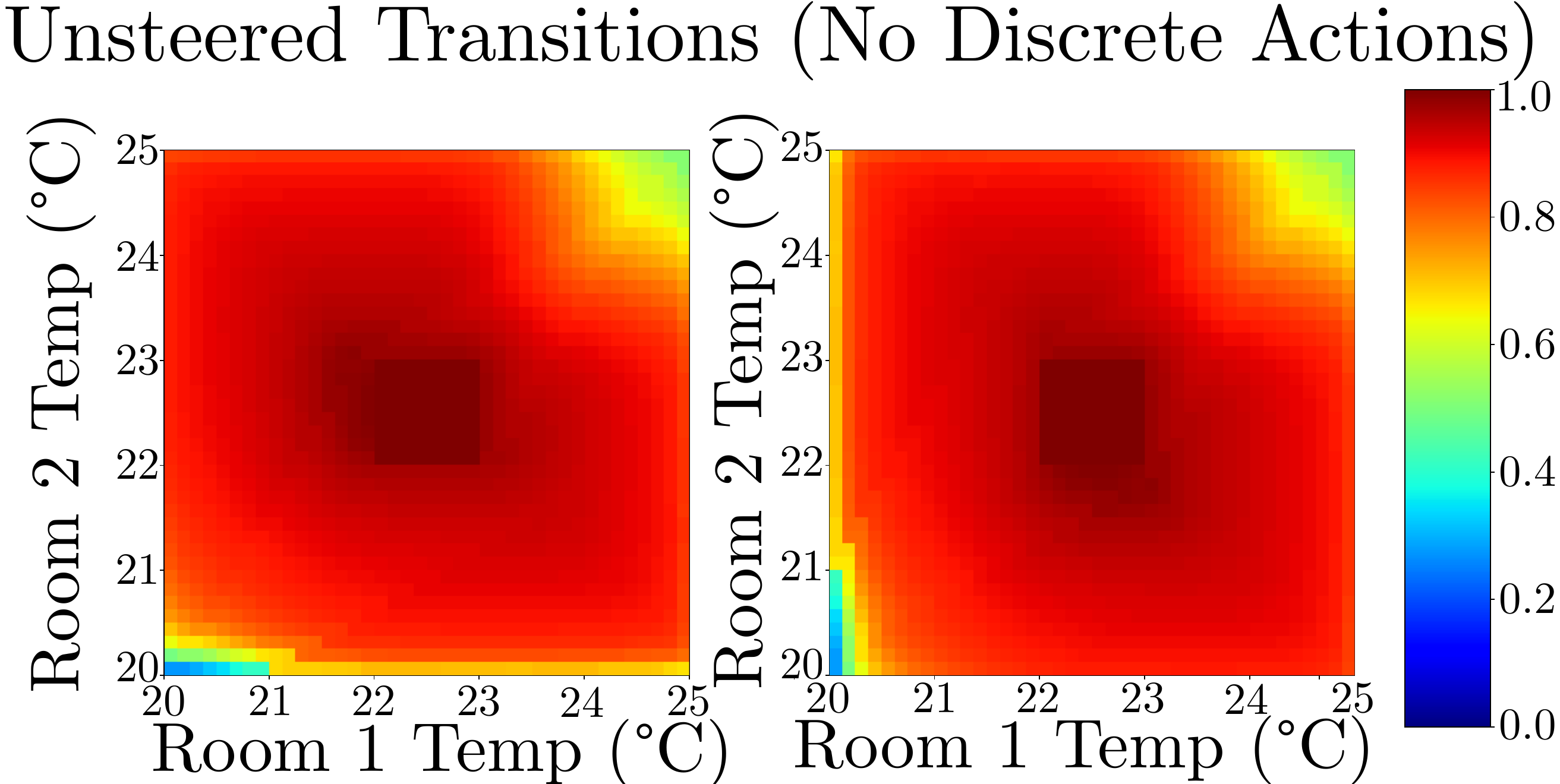}
    \caption{Results for a temperature control problem, with mode transitions that are driven by an MC, with probabilities known to be in the interval 40-60\%.}
    \label{fig:MC}
\end{figure}

\subsubsection{Modelling Switching actions leads to better controllers}
Thus far, we have considered MJLSs with switching actions, e.g., for the temperature regulation problem, we could control moving the fan heater and radiator.
To show that these switching actions improve the quality of the obtained policies (and thus controllers), we perform a variant of the experiment in which we fix the policy of the mode switching MDP a-priori.
As in the original experiment, we are interested in reaching a goal region defined as a temperature of between 22 and 23$^\circ$C in both rooms, whilst avoiding critical temperatures below 20$^\circ$C, or above 25$^\circ$C. 
In the new result, we model our system under \cref{ass:MDP}, but consider mode transitions to happen at random without any control, but with knowledge that all mode transitions are in the interval $[0.4,0.6]$.
By doing so, the Markov jump process that we use to construct our abstraction is in fact a Markov chain.

The results for this experiment with an a-priori fixed Markov jump policy are shown in \cref{fig:MC}.
Compared to \cref{fig:heat_mdp_all}, we observe that not modelling the mode switching actions in the product iMDP leads to an optimal policy with worse satisfaction probabilities, which in turn leads to a feedback controller with worse guarantees.
In other words, modelling the mode switching actions in the product iMDP and synthesising an optimal policy jointly with the individual-mode actions leads to feedback controllers with better guarantees. 
%
%
\bibliographystyle{splncs04}
\bibliography{Bib.bib}

\end{document}